# Politechnika Warszawska

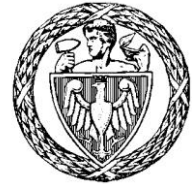

## WYDZIAŁ ELEKTRONIKI I TECHNIK INFORMACYJNYCH

Instytut Informatyki

# Praca dyplomowa inżynierska

na kierunku Informatyka
w specjalności Inżynieria Systemów Informatycznych

Wykorzystanie sztucznej inteligencji
do generowania treści muzycznych

## Mateusz Dorobek

Numer albumu 277285

promotor
prof. dr hab. inż. Przemysław Rokita

konsultacje
mgr inż. Mateusz Modrzejewski

*Niniejszą pracę pragnę zadedykować moim cudownym rodzicom*
***Halinie i Andrzejowi Dorobek**, których nieocenione wsparcie*
*i motywacja pozwoliły mi poświęcić się nauce i zdobywać bezcenne*
*wykształcenie podczas studiów na Politechnice Warszawskiej*

# Streszczenie

**Tytuł**: *Wykorzystanie sztucznej inteligencji do generowania treści muzycznych.*

Praca omawia zagadnienie generowania muzyki z wykorzystaniem głębokich splotowych generatywnych sieci przeciwstawnych (DCGAN). Do nauki sieci zostały użyte muzyczne bazy nagrań MIDI, zawierające muzykę klasyczną i jazzową. Opisany jest algorytm przetwarzania pliku MIDI do obrazu w formacie taśmy pianoli (ang. piano roll) zgodnym z rozmiarem wejściowym sieci, używając kanałów RGB obrazu jako dodatkowych nośników informacji, a także algorytm dekompresji z obrazu do pliku muzycznego. Sieć neuronowa nauczyła się generować obrazy nierozróżnialne od danych wejściowych. Muzyka uzyskana z tak wygenerowanych obrazów charakteryzuje się występowaniem struktur harmonicznych i rytmicznych. Podsumowane są wnioski z kilku przeprowadzonych eksperymentów. Poddane dyskusji zostały możliwe kierunki rozwoju i porównanie z wybranymi istniejącymi rozwiązaniami.

**Słowa kluczowe**: generowanie muzyki, splotowe sieci neuronowe, generatywne sieci przeciwstawne, przetwarzanie obrazów, uczenie głębokie, sztuczna inteligencja, uczenie maszynowe, sieci neuronowe, muzyka, jazz, midi.

# Abstract

**Title**: *Use of artificial intelligence for generating musical contents.*

This thesis is presenting a method for generating short musical phrases using a deep convolutional generative adversarial network (DCGAN). To train neural network were used datasets of classical and jazz music MIDI recordings. Our approach introduces translating the MIDI data into graphical images in a piano roll format suitable for the network input size, using the RGB channels as additional information carriers for improved performance. The network has learned to generate images that are indistinguishable from the input data and, when translated back to MIDI and played back, include several musically interesting rhythmic and harmonic structures. The results of the conducted experiments are described and discussed, with conclusions for further work and a short comparison with selected existing solutions.

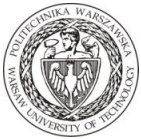 **Politechnika Warszawska**

………........................
miejscowość i data

……………………………..
imię i nazwisko studenta

……………………………..
numer albumu

………………….…………
kierunek studiów

## OŚWIADCZENIE

Świadomy odpowiedzialności karnej za składanie fałszywych zeznań oświadczam, że niniejsza praca dyplomowa została napisana przeze mnie samodzielnie, pod opieką kierującego pracą dyplomową.

Jednocześnie oświadczam, że:

– niniejsza praca dyplomowa nie narusza praw autorskich w rozumieniu ustawy z dnia 4 lutego 1994 roku o prawie autorskim i prawach pokrewnych (Dz.U. z 2006 r. Nr 90, poz. 631 z późn. zm.) oraz dóbr osobistych chronionych prawem cywilnym,

– niniejsza praca dyplomowa nie zawiera danych i informacji, które uzyskałem w sposób niedozwolony,

– niniejsza praca dyplomowa nie była wcześniej podstawą żadnej innej urzędowej procedury związanej z nadawaniem dyplomów lub tytułów zawodowych,

– wszystkie informacje umieszczone w niniejszej pracy, uzyskane ze źródeł pisanych i elektronicznych, zostały udokumentowane w wykazie literatury odpowiednimi odnośnikami,

– znam regulacje prawne Politechniki Warszawskiej w sprawie zarządzania prawami autorskimi i prawami pokrewnymi, prawami własności przemysłowej oraz zasadami komercjalizacji.

Oświadczam, że treść pracy dyplomowej w wersji drukowanej, treść pracy dyplomowej zawartej na nośniku elektronicznym (płycie kompaktowej) oraz treść pracy dyplomowej w module APD systemu USOS są identyczne.

…………………………………….
czytelny podpis studenta

.

# Spis Treści

# 1 Wstęp

Muzyka jest nieodzownym elementem naszego życia, towarzyszy nam na każdym jego etapie. Zarówno w dniu codziennym, jak i w wyjątkowych chwilach odpowiednia muzyka jest bardzo ważna. Historia muzyki zaczęła się intensywnie rozwijać dopiero od końca XVIII w., natomiast jej początki sięgają ponad 3000 lat p.n.e. [1] Muzyka przez setki lat bardzo się zmieniła i przyjmowała bardzo rożne formy, a jej rozwój jest bardzo dynamiczny nawet współcześnie. Pomimo tak wielu różnic pomiędzy różnymi gatunkami i stylami muzycznymi, nadal łączy je wiele wspólnych zasad. Teoria muzyki jest obszerną dziedziną nauki, która bardzo szczegółowo opisuje takie aspekty jak kompozycja, aranżacja, czy harmonia, natomiast praktyka dowodzi, że odchodzenie od tych zasad często daje ciekawe rezultaty. To jedynie potwierdza, jak złożoną dziedziną sztuki jest muzyka oraz jak bliska ona jest człowiekowi. Wszelkie reguły jedynie próbują opisać to, co człowiek stworzył od praktycznie zera. Teoretycznie rzecz ujmując, muzyka jest niczym więcej niż sumą okresowych zakłóceń ośrodka o różnych częstotliwościach. Muzyka, jaka ukształtowała się w naszej kulturze, nie jest jedynie losowym szumem, a stanowi harmoniczną spójną całość, którą można próbować modelować.

Wpływ technologii na muzykę jest ogromny. Jak każda dziedzina sztuki rozwija się pod wpływem wielu czynników zewnętrznych, a kierunku tego rozwoju praktycznie nie sposób przewidzieć. Stworzenie nowych i udoskonalanie dawnych instrumentów, czy wynalezienie płyty winylowej na przełomie XIX i XX wieku, powstanie radia w pierwszej połowie XX wieku, czy rozpowszechnienie się płyt CD w latach osiemdziesiątych. Te wszystkie technologie zrewolucjonizowały muzykę, rozwinął się profesjonalny rynek muzyczny. Wraz z powstaniem Internetu, portali muzycznych, serwisów streamingowych dostępność muzyki osiągnęła ogólnoświatowy poziom. W ułamku sekundy jesteśmy w stanie odsłuchać nagrania z dowolnej części świata, udostępnić swoją twórczość i być na bieżąco z najnowszymi trendami w muzyce. Ma to swoje dobre i złe strony, natomiast niezaprzeczalny jest wpływ technologii na kształt muzyki zwłaszcza współcześnie.

Muzyka jazzowa znacząco wyróżnia się na tle innych gatunków. Przede wszystkim razem z poprzedzającym ją bluesem stanowi fundament całej muzyki rozrywkowej. Rozmaitość stylów, długa historia i całe spektrum aparatu wykonawczego muzyki jazzowej sprawia, że ta muzyka zasługuję na szczególną uwagę. Istotny z punktu widzenia tej pracy jest aspekt złożoności tej muzyki. Z jednej strony nieograniczona niemal swoboda wykonawcza, z drugiej obszerna teoria harmonii jazzowej. To połączenie daje wyjątkowo złożoną strukturę muzyczną, której analiza, czy modelowanie wymaga nawet nie tyle dużej wiedzy ile doświadczenia. Jest to zarazem skomplikowany problem i intrygujące wyzwanie.

## 1.1 Cel i motywacja pracy

Jako aktywny muzyk przez całe życie jestem bardzo blisko z muzyką, która towarzyszy mi na co dzień. Będąc uczniem szkół muzycznych miałem do czynienia z teoretycznym podejściem do tej dziedziny sztuki. Grając z wieloma innymi muzykami w różnorodnych formacjach, doświadczam muzyki od strony praktycznej. Analiza dzieła muzycznego to proces, który zahacza zarówno o sferę nauk technicznych pozwalających zrozumieć naturę dźwięku, fali akustycznej, mechanizm jej generowania przez instrumenty, a także o teorię muzyki, która stara się opisać jej strukturę, do momentu, gdy dotykamy takiej sfery muzyki jak uczucia, które potrafi ona wywołać nastrój, który jej towarzyszy i niezliczoną liczbą wrażeń, których nie można opisać, a można odczuć. W tej pracy zamierzam stworzyć program zdolny do stworzenia treści muzycznych podobnych do utworów stworzonych przez człowieka, a następnie podjąć się oceny tych treści w kontekście przyjętych założeń projektowych.

## 1.2 Założenia projektu

Podstawowym założeniem pracy dyplomowej jest generowania krótkich fraz muzycznych. Powinny one zawierać fragmenty, w których występują typowe dla współczesnej muzyki struktury harmoniczne, powtarzalne frazy rytmiczne. Dodatkowo oczekuję, że generowane frazy będą posiadały wykrywalne zależności między trzema najważniejszymi elementami tworu muzycznego: melodią, harmonią i rytmem. Dodatkowym atutem będzie z pewnością zależność czasowa nie tylko na poziomie rytmu, ale też na poziomie harmonii i melodii. Nie zakładam natomiast, że generowane fragmenty będą stanowiły spójną całość lub będą wolne od błędów. Celem tego projektu jest zbadanie możliwych rozwiązań, jakie oferuje sztuczna inteligencja w dziedzinie generowania muzyki.

## 1.3 Pomoc merytoryczna

Chciałbym serdecznie podziękować Panu mgr inż. Mateuszowi Modrzejewskiemu za wsparcie merytoryczne i konsultacje dotyczące tematyki pracy, możliwych i sprawdzonych rozwiązań, zakresu przeprowadzonych eksperymentów i metod analizy otrzymanych wyników.

## 2 Istniejące rozwiązania w dziedzinie generowania muzyki

W tym rozdziale opiszę część z istniejących rozwiązań dotyczących generowania i przetwarzania muzyki z wykorzystaniem sieci neuronowych. W szczególności zdolne do transkrybowania, komponowania czy tworzenia i łączenia brzmień.

### 2.1 Hexahedria - Summer Research on the HMC

Celem tego projektu [2] było stworzenie oprogramowania zdolnego do generowania solówek jazzowych na podstawie struktury harmonicznej utworu. Bardzo pożądaną cechą architektur stosowanych w tego typu problemach jest niezależność od tonacji, czyli uczenie sieci do generowania solówek w jednej tonacji. Pozwala to wykorzystać tak zdobyte doświadczenie do generowania solówek również w dowolnej tonacji, nawet takiej, która nie występowała w zbiorze uczącym. Tego typu architektury mają wzgląd na harmonię nie tylko jako ciąg niezależnych akordów, lecz także na powiązaną strukturę, gdzie każdy akord pełni jakąś funkcję względem swoich sąsiadów. Oczywiście jest to podejście, które odzwierciedla rzeczywistą strukturę harmonii, która nie jest zależna od tonacji, w której jest utwór. Aby to osiągnąć, trzeba było uzależnić aktualną nutę od jakiegoś punktu odniesienia. Można to zrobić na dwa sposoby: uzależniając aktualną nutę od poprzedniej lub od aktualnego akordu. Obydwa podejścia mają swoje niebywałe zalety, jedna gwarantuje melodyczność, druga zgodność z harmonią utworu. Aby połączyć te dwa podejścia, skorzystano z metody Product-of-Experts.

#### 2.1.1 Model generatywny produktu ekspertów

Produkt ekspertów (ang. Product of Experts) to sposób na połączenie dwóch dystrybucji w jedną, dzięki czemu możemy skutecznie wykorzystać wyniki uzyskane z dwóch różnych sieci.

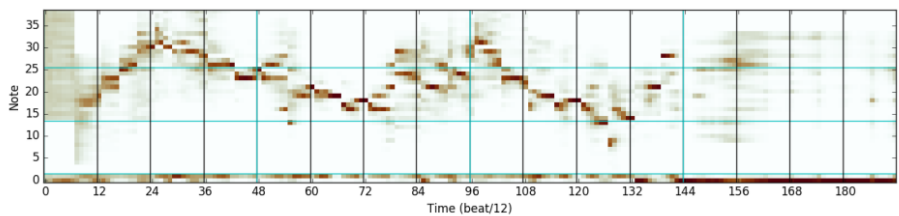

*Rysunek 1 Dystrybucja niezależna, bez wykorzystania metody. [2]*

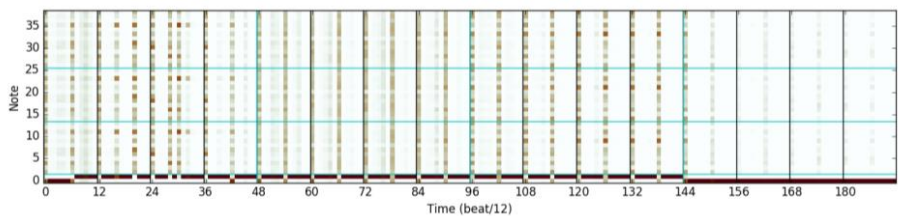

*Rysunek 2 Dystrybucja zależna od bieżącego akordu. [2]*

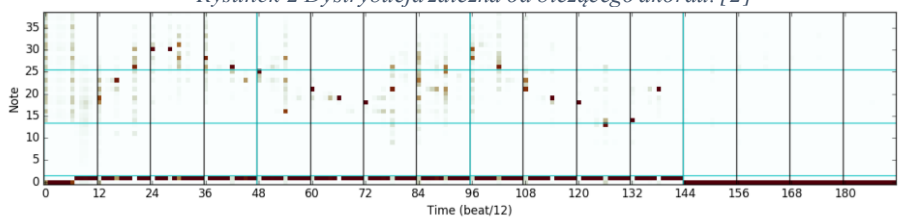

*Rysunek 3 Dystrybucja zależna od poprzedniego dźwięku. [2]*

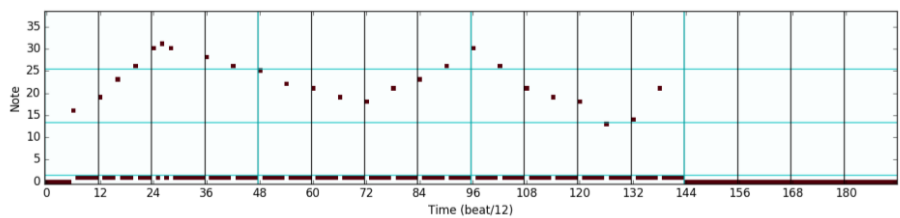

*Rysunek 4 Dystrybucja łączona z wykorzystaniem Product-of-Experts. [2]*

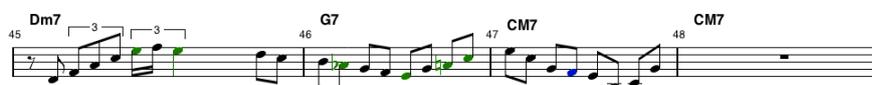

*Rysunek 5 Zapis nutowy dystrybucji łączonej. [2]*

### 2.1.2 Autoenkoder i kompresowanie sekwencji

Istnieje specjalny typ improwizacji nazywany zawołanie i odpowiedź lub inaczej czwórki, gdzie dwóch lub więcej muzyków na przemian gra krótkie improwizowane fragmenty, gdzie każdy jest pewną wariacją poprzedniego. Naiwnym podejściem byłaby lekka modyfikacja wysokości i czasu niektórych dźwięków. Odpowiednim podejściem okazują się autoenkodery, które pozwoliłyby zenkodować fragmenty muzyczne i lekko modyfikować tak skompresowany wektor, dzięki czemu poruszamy się w przestrzeni dobrych solówek, a nie przypadkowo zmodyfikowanych.

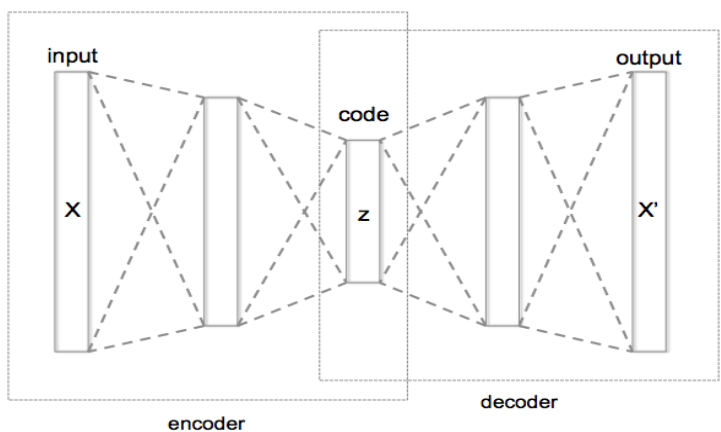

*Rysunek 6 Struktura autoenkodera. [3]*

Dwa główne podejścia:

- Autoenkoder globalny - czyta całą sekwencję, tworzy pojedynczy wektor i dekoduje z powrotem do całej sekwencji. Problemem, jaki występuje przy tym rozwiązaniu jest to, że nie ono może zenkodować melodii o dowolnej długości do wektora o zdefiniowanej długości.
- Autoenkoder typu Timestep-Level [4] - enkoduje pojedynczy odcinek czasowy jako wektor i później go dekoduje i bierze się za kolejny odcinek uzależniony już od wyniku pierwszego. Problemem jest to, że nie potrafi reprezentację sekwencji jako całość.

Rozwiązaniem w tym projekcie okazała się hybryda. Autoenkoder produkuje sekwencje enkodowanych wektorów, ale produkowane w ten sposób sekwencje muszą być krótsze niż sekwencja wejściowa. Na przykład może zwracać pojedynczy wektor dla każdej połówki taktu na wejściu. To pozwala na przetrzymywanie dość długich solówek i jednocześnie uzyskiwanie sekwencji wektorów, które razem mogą reprezentować kolejne części melodii wejściowej.

### 2.1.3  Generatywna sieć przeciwstawna

Podstawowym założeniem generatywnych sieci przeciwstawnych [5] (ang. generative adversarial networks – GAN) jest wykorzystanie dwóch, konkurujących sieci neuronowych:

- Dyskryminatora, który próbuje odróżnić wytrenowane dane od wygenerowanych przez generator.
- Generatora, który generuje nowe dane i próbuje oszukać dyskryminator, żeby oceniał wygenerowane dane jako prawdziwe.

GAN-y najlepiej radzą sobie ze skomplikowanymi wielowymiarowymi danymi oraz z ciągłymi danymi wejściowymi, ale na potrzeby projektu Summer Research on the HMC spróbowano z dyskretnymi sekwencjami, których rozmiar i liczba jest mniejsza niż dla obrazów. Jednym z problemów, jaki się pojawił była propagacja wsteczna wyjścia dyskryminatora. Rozwiązaniem było wzięcie wyjścia dyskryminatora dla konkretnego odcinka czasu i zastosowanie propagacji wstecznej korzystając z dystrybucji generatora dla tego odcinka. Dzięki czemu funkcja dyskryminatora będąca sumą ważoną stała się w pełni różniczkowalna względem współczynników generatora. Kolejnym problemem było to, że w początkowej fazie generator generował losowe nuty, co dyskryminator wykrywał bardzo wcześnie, przez co nauczył się patrzeć jedynie na początek melodii. W efekcie nie był różniczkowalny po całości wejścia, a jedynie po jego początkowym fragmencie.

## 2.2 Projekt Google - Magenta

Magenta [6] to otwarto-źródłowy projekt zbudowany na TensorFlow. Obejmuje głównie dwie dziedziny: muzyka i obraz. Niżej opisze szereg aplikacji, które powstały w oparciu o framework, jak zapewnia Magenta. Do każdego z projektów dodam komentarz odnośnie do potencjału sztucznej inteligencji, jaki został w nim wykorzystany oraz zastosowaniu danego projektu w muzyce. Mówiąc o zastosowaniu technologii w muzyce, mam na myśli jej profesjonalny aspekt, pomijam możliwe zastosowanie w muzyce amatorskiej czy hobbystycznej.

### 2.2.1  Melody Mixer i Latent Loops.

Prosta aplikacja pozwalająca z dwóch/czterech melodii utworzyć zbiór kilkunastu melodii zaczynający się od jednej a kończący na drugiej melodii, gdzie każdy z kolejnych fragmentów staje się coraz bardziej podobny do końcowej melodii. Dostępna jest w wersji online [7] [8]. Program pokazuje możliwości AI w dziedzinie kompresowania dekodowania i enkodowania muzyki przez sieci neuronowe oraz umieszczenia jej fragmentów w wektorowej przestrzeni, dzięki czemu możemy tak swobodnie się przemieszczać między fragmentami muzycznymi. Przydatność dla profesjonalnych zastosowań muzycznych nie jest natomiast zbyt duża. Można wykorzystać płynne zmiany melodii podczas przejść w muzyce elektronicznej, rozłożonych akordach czy arpeggiach.

### 2.2.2 Beat Blender

Beat Blender, podobnie jak Latent Loops, pozwala na płynne przechodzenie pomiędzy podanymi na wejściu, ale rytmami, a nie melodiami. Tutaj również jak w Latent Loops zastosowanie w muzyce elektronicznej i raczej nie do odgrywania na żywo, lecz w produkcji. Można przetestować Beat Blender w wersji online [9].

### 2.2.3 NSynth

Aplikacja, która jest wynikiem projektu [10] służy do syntezy brzmień na podstawie sampli podanych na wejście. W przeciwieństwie do standardowych narzędzi syntezy takich jak oscylatory, synteza FM (ang. Frequency Modulation), PCM (ang. Pulse-Code Modulation), czy synteza tablicowa NSynth używa sieci neuronowych do syntezy na poziomie pojedynczych sampli, dając możliwość sterowania takimi (wysokopoziomowymi) parametrami, jak tembr i dynamika. Przy procesie uczenia wykorzystana była ponad 300 tysięczna baza dźwięków [11] dostarczona również przez Magentę. Zastosowanie tego programu to przede wszystkim muzyka elektroniczna, czyli brzmienia do instrumentów klawiszowych oraz postprodukcja.

### 2.2.4 Performance RNN

Jeden z ciekawszych projektów w tym zestawieniu. Aplikacja służy do generowania muzyki, która przypomina akompaniament fortepianu. Prosty interfejs pozwalający na wprowadzenie wag na odpowiednie składniki skali chromatycznej pozwala uzyskać odpowiedni nastrój w zależności od funkcji harmonicznej, którą przedstawiają wagi na kolejnych dźwiękach. Charakter tego akompaniamentu przypomina ten, który można usłyszeć w muzyce klasycznej. Wynik przypomina stylistycznie grę Chopina, Debussy'ego, Mozarta. Aktualnie istnieją na rynku o wiele lepsze programy typu automatyczny akompaniator, ze znacznie większą różnorodnością stylów i możliwością dokładnego odegrania harmonii utworu, lecz Performance RNN [12] jako jeden z niewielu jest zbudowany na sieci neuronowej. Na razie nie stanowi dla nich poważnej konkurencji, lecz ma potencjał, który sprawia, że warto śledzić ten i podobne mu projekty.

### 2.2.5 AI Duets

Aplikacja [13] ta potrafi naśladować/odpowiadać na melodię zagna na klawiaturze lub sterowniku MIDI. Wstępne wnioski:

- Posiada świadomość tonacji, w jakiej się znajduje.
- Posiada świadomość, jaka jest gęstość rytmu, rozpiętość melodii i reaguje na to.
- Nie potrafi grać harmonicznie, jedynie monofonicznie.
- Czasami cytuje bezpośrednio.
- Nie potrafi grać w tempie.

Na chwilę obecną nie ma żadnego profesjonalnego zastosowania dla tego typu aplikacji, lecz jest to dobry punkt wyjścia do oprogramowania zdolnego do grania na żywo improwizacji typu zawołanie i odpowiedź.

### 2.2.6  Neural Melody Autocompletion

Efekt działania podobny jak w A.I. Duets, aplikacja będąca rozszerzeniem wytrenowanej sieci ImprovRNN. Udało się również uzyskać świadomość harmonii, lecz melodie nadal są arytmiczne, i donikąd nie dążą. Wejściem jest akord grany na klawiaturze, lecz przy tak płytkiej (jeden akord) świadomości harmonii nie ma mowy o złożonej harmonii. Reakcja na zmianę harmonii też nie jest zbyt szybka ok. 2 sekund, więc na scenie okaże się bezużyteczny. Efekt działania aplikacji można odsłuchać na witrynie internetowej [14].

### 2.2.7  Onsets and Frames

Zdecydowanie najciekawszy projekt [15] który bazuje na bibliotece, jaką oferuje Magenta. Program potrafi transkrybować (tworzyć zapis nutowy utworu na podstawie zapisu audio) polifoniczne nagranie fortepianu, osiągając przy tym bardzo obiecujące wyniki. Wewnątrz posiada sieci splotowe oraz LSTM. Autorzy za przyczynę tak dużej skuteczności uznają podział detekcji nut na dwa stosy sieci neuronowych. Jeden stos (Onset) wykrywa pierwsze kilka klatek, w których nuta jest grana, a drugi stos (Frame) wykrywa te klatki, w których dźwięk jest nieaktywny. Surowe dane z części wykrywające początki nut są przekazywane jako dodatkowy argument do drugiego stosu.

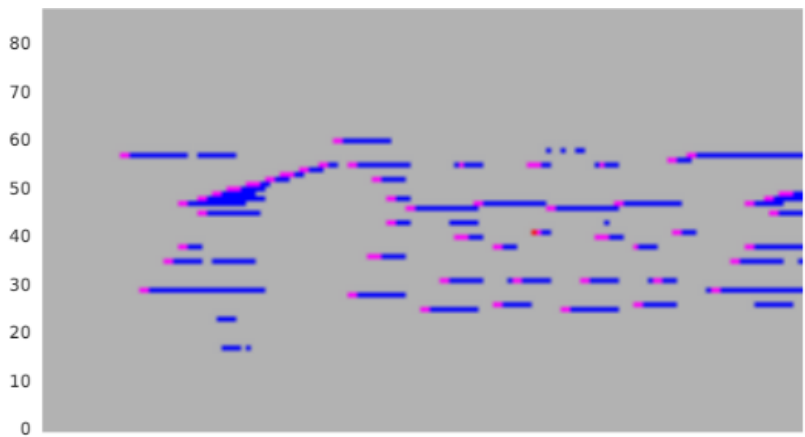

*Rysunek 7 Predykcje stosu Frames. [15]*

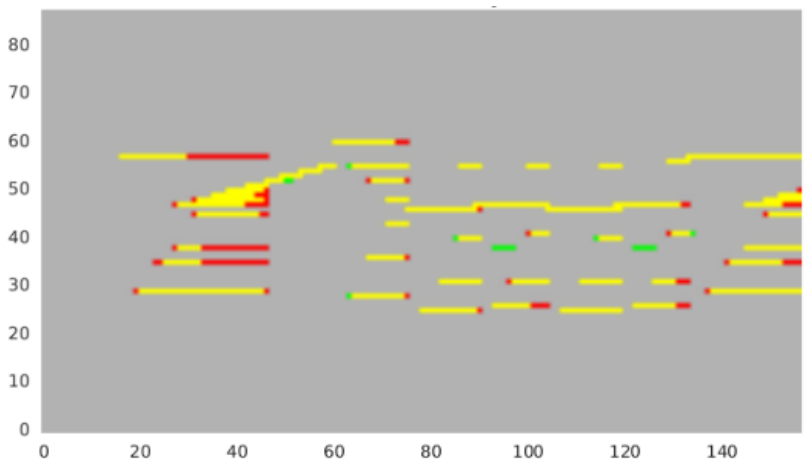

*Rysunek 8 Predykcje stosu Frames po korekcji przez predykcje stosu Onset. [15]*

Nuty, które zaczynają się purpurowym kolorem (zob. Rysunek 7) znalazły potwierdzenie od stosu Onset, lecz nie wszystkie i to właśnie te, które nie uzyskały tego potwierdzenia, okazały się fałszywie wykryte, co potwierdza istotę tej drugiej warstwy. (zob. Rysunek 8) Pokazuje na czerwono miejsca, w których stos Frames nic nie wykrył, a na zielono te, które wskazał błędnie. Żółte nuty to zgodność w obu stosach.

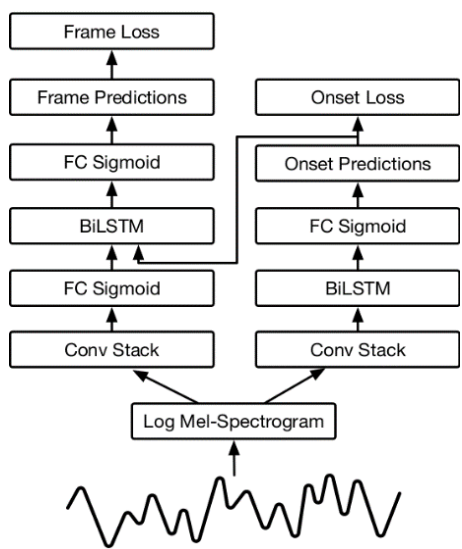

*Rysunek 9 Struktura modelu Onsets and Frames. [15]*

## 2.3 Amper Music

Amper Music to komercyjny projekt [16], który oferuje generowanie fragmentów muzyczne o bardzo różnej stylistyce. Pozwala manipulować takimi parametrami jak instrumentarium, tempo, tonacja, charakter i dynamika utworu. Wygenerowana muzyka jest dość krótka, ale dobrze oddaje zadany przez użytkownika nastrój, tonację i tempo. Dodatkowo Amper korzysta z bardzo dobrych sampli audio do generowania muzyki oraz stosuje dba o bardzo dobrą produkcję wygenerowanej muzyki (dobrze dobrane efekty audio, miks i mastering). Utwory tworzone przez Amper są wykorzystywane w reklamach telewizyjnych. Za około 200 $ można wykupić prawa do wygenerowanego utworu.

## 2.4 GRUV

Projekt Badawczy [17] napisany w Pythonie z wykorzystaniem Kerasa i Theano, powstały w 2015, którego efektem było wygenerowanie bardzo zakłóconego audio, w którym było słychać rytm harmonię, a nawet przebijający się wokal. Struktura, na jakiej się opiera to LSTM i CNN. Projekt nie jest już rozwijany. Naukowcy, którzy się nim zajmowali, trenowali go na małym zbiorze i przeprowadzili jedynie 2000 iteracji. Wyniki tego projektu są bardzo obiecujące dla dziedziny generowania muzyki w formacie audio.

## 2.5 DeepJazz

Projekt [18] o architekturze LSTM korzystający z bibliotek: TensorFlow, Theano i Keras. Model był trenowany na pojedynczym pliku MIDI. Po 128 cyklach treningowych na muzyce Pata Metheny'ego uzyskano rytmiczny utwór o złożonej harmonii z melodią, która korespondowała z akordowym akompaniamentem, lecz był on bardzo podobny do danych wejściowych, doszło tu do zjawiska zbytniego dopasowania (ang. overfitting).

# 3 Sieci neuronowe w zastosowaniach generowania muzyki

W tym rozdziale opisane jest kilka architektur, który mogą znaleźć zastosowanie w tematyce, jaką porusza moja praca inżynierska, czyli w generowaniu treści muzycznych.

## 3.1 Rekurencyjne sieci neuronowe – RNN

Rekurencyjne sieci neuronowe [19] w odróżnieniu od standardowych sieci neuronowych mają zdolność zapamiętania kontekstu, w którym się znajdują. Wiele zagadnień takich jak generowanie muzyki, rozpoznawanie tekstu, modelowanie języka, czy przetwarzanie obrazów, są zależne od poprzedniego stanu. Taką zależność próbują odtworzyć sieci rekurencyjne. Aby tego dokonać, w ich strukturze występują pętle.

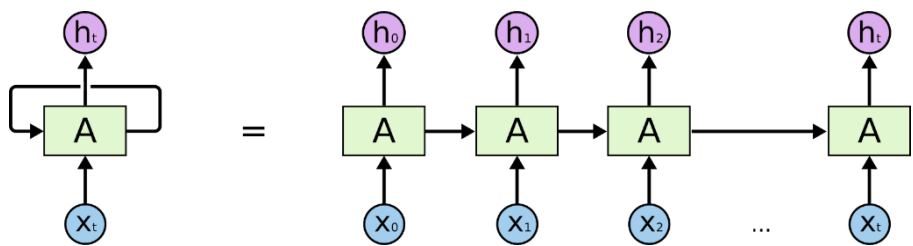

*Rysunek 10 Struktura sieci rekurencyjnych. [19]*

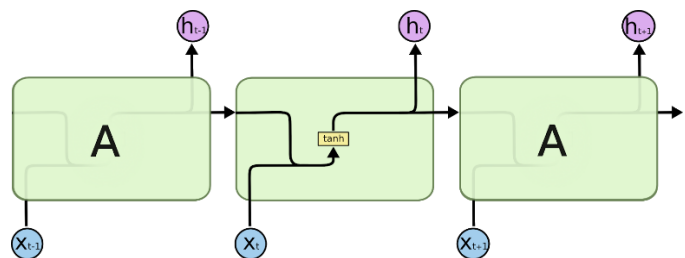

*Rysunek 11 Funkcja aktywacji tanh – tangens hiperboliczny. [19]*

Oznacza to, że pewne fragmenty sieci się powtarzają. Wiele problemów wymaga od naszej sieci zapamiętania krótkiego kontekstu, (zob. Rysunek 12) np. szacowanie prawdopodobieństwa wystąpienia danego słowa po danej sekwencji. Z częścią z nich RNN radzą sobie bardzo dobrze.

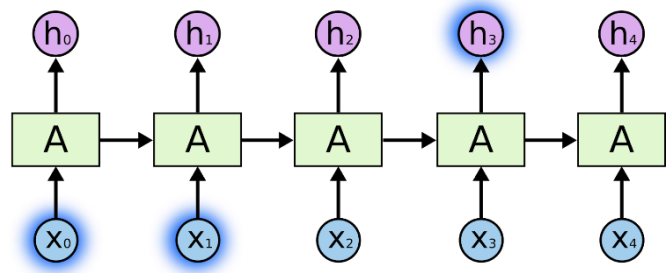

*Rysunek 12 Pamięć krótkotrwała w RNN. [19]*

Problem pojawiać się, gdy dane potrzebują znacznie szerszego kontekstu, (zob. Rysunek 13), aby rozwiązać problem. Teoretycznie standardowe sieci rekurencyjne (Vanilla RNN) mogą się tego nauczyć, ale praktyka tego nie potwierdza.

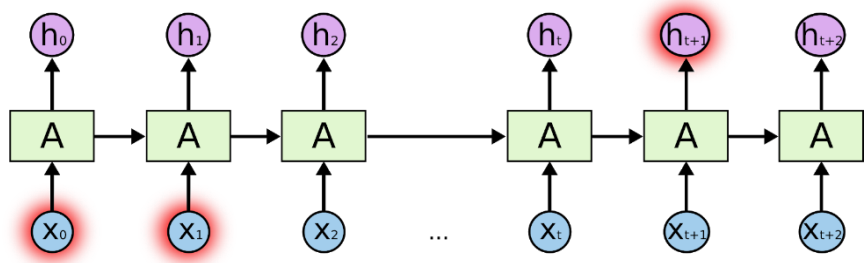

*Rysunek 13 Brak pamięci dla szerszego kontekstu. [19]*

Aby poradzić sobie z tym problemem powstały sieci LSTM, które o wiele lepiej uczą się wykorzystywać kontekst z nawet bardzo odległej przeszłości.

## 3.2 Long Short Term Memory

LSTM to specjalny rodzaj RNN, zdolny do uczenia się długo-terminowych zależności. Podobnie jak klasyczny RNN ma strukturę składającą się z powtarzających się elementów.

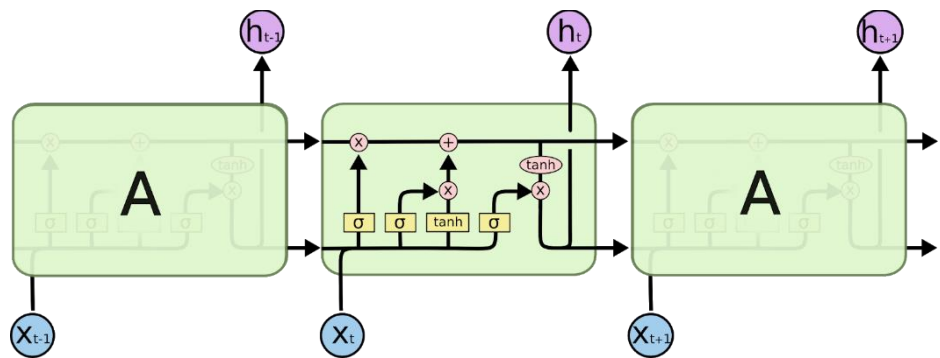

*Rysunek 14 Struktura LSTM. [19]*

### 3.2.1   Budowa podstawowego modelu LSTM

- **Forget gate layer** (zob. Rysunek 15) na tym poziomie decydujemy które informacje będą używane i w jakim stopniu, bramka nadaje każdej wartości na wejściu pewną wagę od 0 do 1.

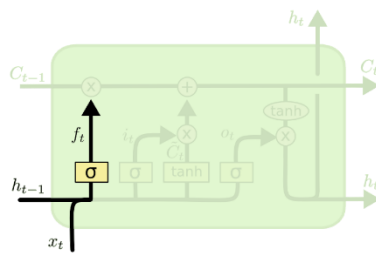

$$f_t = \sigma \left( W_f \cdot [h_{t-1}, x_t] \; + \; b_f \right)$$

*Rysunek 15 Forget gate layer. [19]*

**Input gate layer** (zob. Rysunek 16) - w tym miejscu (warstwa sigmoid-owa) podejmuje się decyzje, które wartości będą aktualizowane. Warstwa tanh tworzy wektor wartości, które są dodane do stanu.

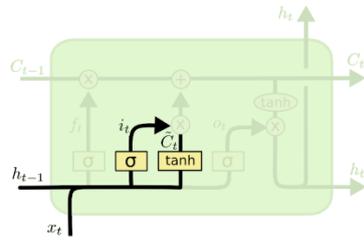

$$i_t = \sigma\left(W_i \cdot [h_{t-1}, x_t] \ + \ b_i\right)$$
$$\tilde{C}_t = \tanh(W_C \cdot [h_{t-1}, x_t] \ + \ b_C)$$

*Rysunek 16 Input gate layer. [19]*

- Aktualizujemy stan naszej komórki (zob. Rysunek 17).

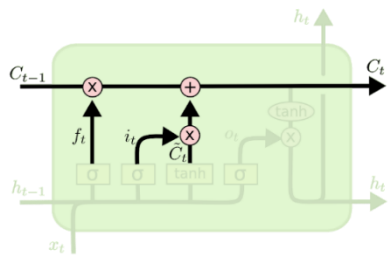

$$C_t = f_t * C_{t-1} + i_t * \tilde{C}_t$$

*Rysunek 17 Operacje macierzowe. [19]*

- **Output gate layer** (zob. Rysunek 18) - na końcu decydujemy które wartości będziemy dawać na wyjściu komórki.

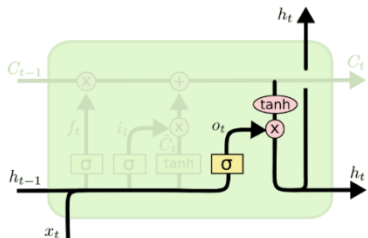

$$o_t = \sigma\left(W_o \ [h_{t-1}, x_t] \ + \ b_o\right)$$
$$h_t = o_t * \tanh\left(C_t\right)$$

*Rysunek 18 Output gate layer. [19]*

### 3.2.2 Warianty LSTM

Przedstawiony wyżej model jest najbardziej podstawowym, istnieją natomiast zmodyfikowane odmiany, o których należy tu wspomnieć. Wersja z dodanym tzw. „peephole connections" (zob. Rysunek 19) pozwalamy niektórym bramkom na wgląd do stanu komórki.

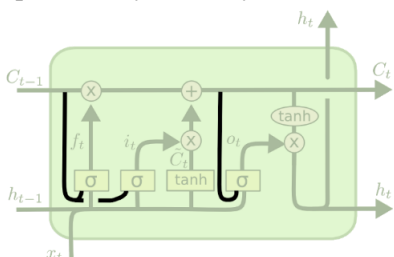

$$f_t = \sigma\left(W_f \cdot [\boldsymbol{C_{t-1}}, h_{t-1}, x_t] \ + \ b_f\right)$$
$$i_t = \sigma\left(W_i \cdot [\boldsymbol{C_{t-1}}, h_{t-1}, x_t] \ + \ b_i\right)$$
$$o_t = \sigma\left(W_o \cdot [\boldsymbol{C_t}, h_{t-1}, x_t] \ + \ b_o\right)$$

*Rysunek 19 Peephole connections. [19]*

Połączone bramki forget i input. Tu poza pewnym uproszczeniem struktury pojawia się ciekawa zależność, a mianowicie możemy nauczyć się nowej rzeczy jedynie, gdy stara jest zapominana.

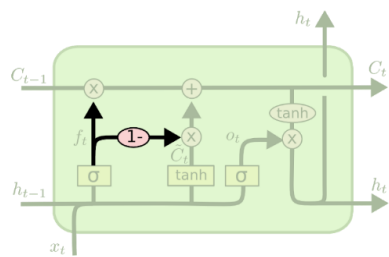

$$C_t = f_t * C_{t-1} + (1 - f_t) * \tilde{C}_t$$

*Rysunek 20 Zapominanie starych zależności [19]*

### 3.2.3 Gated Recurent Unit

Gated Recurent Unit (zob. Rysunek 21) – wersja LSTM, w której bramki forget i input są łączone w jedną bramkę update. Dodatkowo połączony jest stan komórki oraz jej stan ukryty. Występuje kilka innych połączeń wewnątrz struktury niż w standardowym LSTM. Rezultatem jest prostszy model, który jest ostatnio coraz bardziej popularny.

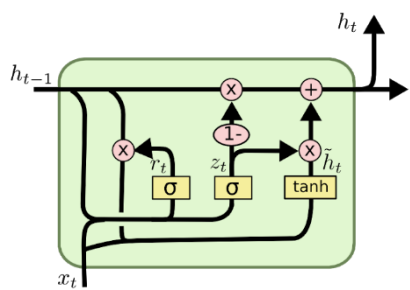

$$z_t = \sigma \left( W_z \cdot [h_{t-1}, x_t] \right)$$
$$r_t = \sigma \left( W_r \cdot [h_{t-1}, x_t] \right)$$
$$\tilde{h}_t = \tanh \left( W \cdot [r_t * h_{t-1}, x_t] \right)$$
$$h_t = (1 - z_t) * h_{t-1} + z_t * \tilde{h}_t$$

*Rysunek 21 Struktura Gated Recurent Unit. [19]*

Zalety sieci LSTM w dziedzinie generowania muzyki:

- Świetnie generuje tekst, więc muzyka w formacie tekstowym MIDI.
- Szybko się uczy (w porównaniu do sieci rekurencyjnych).
- Może generować dane bez końca, co oznacza nie istnieje górne ograniczenie długości muzyki.
- Łatwo uczy się relacji przyczyna-skutek, co w muzyce pełni dużą rolę. Oprócz struktury pionowej - harmonia, posiada ona zależność od czasu, po dominancie następuje rozwiązania, a po idącej w górę melodii kierunek zmienia się w pewnym momencie na odwrotny.
- Dyskretna reprezentacja danych. Każda nuta ma konkretną wysokość, strój jest ustalony na konkretne wartości.

Wady

- Musi przenieść format 2D na 1D - zniekształca się obraz tego, czym jest muzyka.
- Przez pamięć, którą posiada dochodzi do niekorzystnego zjawiska eksplozji gradientu (ang. gradient explosion).

### 3.2.4 Attention - Kolejny krok w rozwoju RNN i LSTM

Mechanizm, opisany w artykule [20], który pozwala obejść problem stałej długości wewnętrznej reprezentacji wektora danych w architekturze Enkoder-Dekoder w szczególności w przypadku gdy dane wejściowe są dłuższe niż dane treningowe. Jest to architektura, która działa na dwóch sieciach LSTM, z których jedna przekształca sekwencję wejściową na wektor o stałej długości, a druga go dekoduje. Mechanizm uwagi (ang. Attention) pozwala uwolnić się od problemu stałej długości wewnętrznej struktury. Osiągane jest to przez przechowywanie uśrednionych wyjść z enkodera LSTM dla każdego kawałka sekwencji wejściowej i trenowanie modelu tak, żeby zwracał selektywną uwagę na te wartości wejściowe i uzależniał od nich wartości sekwencji wyjściowej. Zwiększa to złożoność obliczeniową modelu, ale skutkuje to skuteczniejszym jego działaniem.

Podobna technika jest stosowana w przypadku dużych obrazów i nazywa się ją glimps-based modification ( modyfikacja oparta na przebłyskach) polega na tym, że zanim zostanie wykonana finalna predykcja dokonuje się serii mniejszych pomiarów skupionych na mniejszych fragmentach, które później nakierują na poprawną odpowiedź.

## 3.3 Generatywne sieci przeciwstawne - GAN

Istotą generatywnych sieci przestawnych (ang. generative adversarial networks - GAN) [5] są dwa konkurujące modele sieci. Jedna z nich to generator próbujący naśladować prawdziwe dane. Druga to dyskryminator otrzymuje dane zarówno te prawdziwe, jak i wytwarzane przez generator i próbuje ocenić ich prawdziwość. Przykładowa implementacja: HyperGAN [21]. Zalety sieci typu GAN w dziedzinie generowania muzyki:

- Wyniki w dziedzinie generowania obrazów są bardzo obiecujące.
- Dobrze radzi sobie z danymi przestrzennymi - oznacza to że jest lepiej przystosowany do struktury wejściowej, jaką jest muzyka.
- HyperGAN, wykonując na GPU równoległe obliczenia jest bardzo szybki.

Wady

- Nie potrafi nauczyć się "co jest przed, a co po" - co jest dużą wadą jeśli chodzi o charakter muzyki.
- Nie działa na dyskretnych danych - muzykę można kwantyzować, ponieważ w tekstowym lub graficznym formacie zapisu jest dyskretna, nuty się zaczynają i kończą w konkretnych miejscach i mają konkretną wysokość, która należy do niedużego zbioru.
- Spektrum kolorów pikseli jest szerokie, więc obraz będzie zamazany - w związku z powyższym punktem, jeżeli wyjście jest niedyskretne a wejście dyskretne muzyczny obraz, jaki uzyskamy będzie zamazany.
- Proces uczenia, może się odwrócić - generator i deskryptor mają niechlubną cechę która w pewnych momentach fazy uczenia może spowodować odwrócenie tego procesu.

## 3.4 Splotowe sieci neuronowe

Są to sieci neuronowe najczęściej wykorzystywane w rozpoznawaniu obrazów. Ważną ich cechą jest niezależność od przesunięcia i od przestrzeni, co oznacza, że nie dla lekko innych danych nie będzie konieczne ponowne uczenie się sieci. Ich struktura wymusza wykształcenie swego rodzaju filtrów, które analizują obraz. Przykładowa implementacja: PixelCNN [22] Zaletą sieci splotowych w dziedzinie generowania muzyki jest to, że uczą się kierunkowo i czytają dane wejściowe jako sekwencję, co jest zgodne z charakterem tych danych, ponieważ muzyka ma charakterystykę kierunkową, rozchodzi się w czasie w jednym kierunku.

Te sieci wykorzystuje się, aby uzupełnić brakujące części obrazów, ponieważ generują one dalszą część danych na podstawie tego, co już istnieje, więc w muzyce miałyby zastosowanie na przykład w generowaniu muzyki na żywo. Wadą sieci splotowych jest to, że działają bardzo wolno, ponieważ nic nie może zostać wyprodukowane, póki bezpośredni poprzednik (piksel) nie został stworzony.

## 3.5 Auto-Encoder

Sieć neuronowa wykorzystywana w nienadzorowanym uczeniu maszynowym. Jest to algorytm kompresji danych, który jest dostosowany do danych, na których operuje. Składa się z 3 elementów: kodera, dekodera, funkcji straty. Koder, dekoder i parametryczne funkcje (sieci neuronowe) muszą być różniczkowalne względem funkcji straty będą optymalizowane, aby zmniejszyć stratę wynikające z przeprowadzanej przez nie rekonstrukcji za pomocą metody propagacji wstecznej gradientu, dzięki temu sieć uczy się rozkładać dane wejściowe oraz ignorować szum informacyjny w nich zawarty. Najprostszą implementacją jest Fully-connected Auto Encoder. Znacznie większą popularnością cieszy się VAE – Variational Auto Encoder [23].

# 4 Biblioteki programistyczne używane w uczeniu maszynowym

Implementacja sieci neuronowych to złożone zadania z pomocą przychodzą pomocne narzędzia, z których większość jest darmowa, dobrze udokumentowana z setkami przykładów dostępnych w Internecie. Niektóre z nich posiadają wysokopoziomowe API pozwalające bardzo szybko implementować i modyfikować sieci neuronowych.

## 4.1 TensorFlow

Stworzona w 2015 roku przez Google Brain Team [24] biblioteka do uczenia maszynowego i głębokich sieci neuronowych. Cechuje się dobrą wydajnością i skalowalnością. W celu zwiększenia wydajności napisano ją w C++ oraz stworzono API w Pythonie. Ponadto biblioteka posiada dobrej jakości dokumentację i skuteczne narzędzie do wyświetlania i analizowania diagramów przepływu danych opisujących wykonywane obliczenia – TensorBoard [25]. Istnieją także wyższe warstwy API takie jak Keras, TF-Slim, czy Estimators, które pozwalają testować i porównywać modele.

TensorFlow [26] można uruchomić na karcie graficznej, co jest jedna z ważniejszych jego go cech, ponieważ obliczenia dotyczące głębokich sieci neuronowych przyspieszają zwykle o jeden rząd wielkości na układach GPU (w szczególności na architekturze obliczeniowej CUDA [27] od firmy NVidia o ogólnym zastosowaniu, w przypadku większości bibliotek) w porównaniu do układów CPU. Prostsze metody maszynowego uczenia się zasadniczo nie wymagają przyspieszenia oferowanego przez układy GPU. Poniżej porównanie trzech popularnych API do TensorFlow.

## 4.2 Keras

Keras [28] to wysokopoziomowe API do TensorFlow [26]. Łatwo zacząć z nim pracę, ale jest mniej wydajny niż TFlearn, czy TensorLayer. Oryginalnie był stworzony do pracy z Theano, więc tam posiada lepsze wsparcie, ale po ogromnym sukcesie TensorFlow, również do niego zostało stworzone. Kerasa można obsługiwać bez znajomości TF co jest dużą zaletą, ponieważ w TFlearn i TensorLayer trzeba wykazać się pewną sprawnością w posługiwaniu się TensorFlow. Na początku 2017 roku Google wybrało Kerasa jako domyślne API TensorFlow.

## 4.3 PyTorch

PyTorch [29] jest to stosunkowo młoda biblioteka, który pozwala na dość dużą swobodę implementacyjną, elastyczność, dynamicznie budowany model oraz wbudowany debuger. Coraz więcej osób zaczyna sięgać do niego ze względu na intensywnie rozwijaną dokumentacje i kolejne implementacje najnowszych architektur sieci.

## 4.4 Scikit-learn

Biblioteka Scikit-learn [32] jest napisana w języku Python. Oferuje szeroki wybór skutecznych algorytmów uczenia maszynowego. Dla użytkowników języka Python Scikit-learn to z pewnością najlepsze rozwiązanie spośród wszystkich prostych biblio-tek. Scikit-learn to mocna i sprawdzona biblioteka uczenia maszynowego napisana w języku Python z szerokim asortymentem ugruntowanych algorytmów i zintegrowanych grafik. Program jest stosunkowo łatwy w instalacji, nauce i obsłudze oraz dysponuje dobrymi przykładami i dobrej jakości samouczkami. Dodatkowo, biblioteka Scikit-Learn nie obejmuje implementacji uczenia głębokiego ani uczenia przez wzmacnianie, nie oferuje modeli graficznych ani możliwości prognozowania sekwencji, nie może być też stosowany w przypadku języków innych niż Python. Nie sprawia w rzeczywistości żadnych problemów, jeśli chodzi o szybkość uczenia się. Korzysta z kompilera Cython (Python-to-C) dla funkcji, które muszą być wykonywane błyskawicznie, takich jak pętle wewnętrzne.

## 4.5 Caffe

Caffe [33] to biblioteka do implementacji głębokiego uczenia, która początkowo służyła jako narzędzie do klasyfikacji obrazów. Rozwój Caffe nieco zwolnił, biorąc pod uwagę nieustannie pojawiające się błędy, zatrzymanie rozwoju na wersji 1.0 sprzed ponad roku i opuszczenie projektu przez jego inicjatorów. Framework ten nadal jednak dysponuje dobrej jakości splotowymi sieciami neuronowymi do rozpoznawania obrazów i dobrym wsparciem dla układów GPU o nazwie CUDA od firmy NVidia, a także prostym formatem opisu sieci. Z drugiej strony, Caffe często wymagają znacznych ilości pamięci GPU (ponad 1 GB) do ich uruchomienia, jego dokumentacja nadal zawiera błędy i jest problematyczna, trudno uzyskać wsparcie, a proces instalacji budzi wątpliwości, szczególnie w zakresie obsługi notebooka Python.

## 4.6 Theano

Theano [34] to biblioteka konkurencyjna dla TensorFlow, wyszła w 2016 ostatnia aktualizacja do wersji 1.0.0. w listopadzie 2017. Narzędzie wydaje się bardziej niskopoziomowe niż TF, możliwe, że nawet bardziej wydajne, ale w porównaniu do konkurenta ma mniejsze wsparcie i nie jest tak intensywnie rozwijane.

## 4.7 Microsoft Azure ML Studio

Jest to narzędzie do współpracy, obsługiwane metodą „przeciągnij i upuść", które służy do budowania, testowania i wdrażania rozwiązań z zakresu analizy predykcyjnej na podstawie posiadanych danych. Usługa Machine Learning Studio [35] publikuje modele jako usługi sieci Web, które mogą być łatwo używane w niestandardowych aplikacjach albo narzędziach do analiz biznesowych. Machine Learning Studio to połączenie analiz danych, analiz predykcyjnych, zasobów w chmurze oraz samych danych. W ramach darmowego konta można uczyć sieć w ich chmurze, przez pewien ograniczony czas w miesiącu, posiadają graficzne API, w którym taką sieć można zbudować.

## 4.8 Google Colaboratory

Google Colaboratory [36] jest to rozwiązanie, które zapewnia bardzo szybki start, ponieważ nie wymaga żadnego dodatkowego uwierzytelniania poza kontem google ( AWS [37] GCP [38] Azure [35] wymagają uwierzytelnienia legitymacji studenckiej i podania danych do karty kredytowej). Wirtualna maszyna, jaką oferuje GC posiada system Linux, Kartę graficzną NVidia Tesla K80 (12GB RAM) co jest bardzo potężnym narzędziem w dziedzinie Machine Learningu. Niestety maksymalny czas ciągłej pracy to tylko 12 godzin więc wymagało to ode mnie zapisywania parametrów modelu i optymalizatora oraz ponownego stawiania całego środowiska po upływie tego czasu.

# 5  Dane uczące dla sieci neuronowej

Dane treningowe, jakich można użyć do uczenia sieci neuronowych, muszą mieć finalnie postać tensora liczbowego. Poniżej wypisałem popularne formaty muzyczne, jakich można by było użyć do wygenerowania takich tensorów poprzez zastosowanie odpowiednich konwersji, kompresji i transformacji w procesie preprocesingu.

## 5.1 Format Danych

Muzykę poza jej oczywistą formą, czyli dźwiękową można reprezentować także w postaci zapisu nutowego. Podobnie jest w świecie cyfrowym, gdzie również występują tego typu reprezentacje.

### 5.1.1  Audio

- MP3 — format audio, który powstaje w wyniku kompresji stratnej, uwzględnia model psychoakustyczny człowieka, przez co widmo charakterystyki częstotliwościowej ulega zmianie.
- WAV — gwarantuje wyższą jakość, nie stosuje się w nim algorytmów kompresji.

### 5.1.2  Tekstowe

- MIDI [39] – format, w którym podstawowymi elementami są zdarzenia wciśnięcia klawisza oraz jego odpuszczenia. Dodatkowo występują zdarzenia takie jak wciśnięcie pedała sustain, zmiana położenia pokrętła bend i wiele innych opisujących zmiany parametrów kontrolerów MIDI. Jest to jeden z najbardziej popularnych formatów cyfrowego zapisu nutowego.

```
MFile 1 2 384                    MFile 1 4 120
MTrk                             MTrk
0 Tempo 500000                   0 Meta SeqName "I've Grown Accustomed"
0 TimeSig 4/4 24 8               0 Meta Text "Doug McKenzie"
1 Meta TrkEnd                    0 TimeSig 4/4 24 8
TrkEnd                           0 KeySig -3 major
MTrk                             0 Tempo 451128
0 PrCh ch=1 p=0                  0 Meta Marker "I've Grown Accustomed"
0 Par ch=1 c=64 v=0             0 Meta TrkEnd
808 On ch=1 n=79 v=89           TrkEnd
1019 On ch=1 n=74 v=77          MTrk
1042 On ch=1 n=79 v=0           0 Meta 0x21 00
1078 On ch=1 n=71 v=77          0 Meta TrkName "Accoustic Piano"
1094 On ch=1 n=74 v=0           0 PrCh ch=1 p=0
1140 On ch=1 n=71 v=0           107 On ch=1 n=63 v=91
1157 On ch=1 n=67 v=71          112 On ch=1 n=51 v=74
1194 On ch=1 n=67 v=0           208 On ch=1 n=65 v=84
```

*Rysunek 22 Przykład pliku MIDI w formie tekstowej*

- ABC [40] – format o bardzo prostej i czytelnej składni. Ma zastosowanie bardziej edukacyjne niż profesjonalne. Poniżej (zob. Rysunek 24) wynik kompilacji kodu (zob. Rysunek 23) ABC.

```
X:1
T:Chords
M:2/4
K:C
[CEGc] [C2G2] [CE][DF]|[D2F2][EG][FA] [A4d4]|]
```

*Rysunek 23 Fragment pliku w notacji ABC [40]*

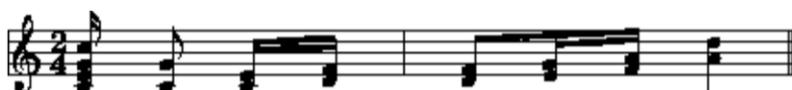

*Rysunek 24 Zapis nutowy - wynik kompilacji pliku ABC (zob. Rysunek 22) [40]*

- MXML – uniwersalny format sposób zapisu plików muzycznych przez edytory nut takie, jak MuseScore, Finale czy Sibelius. Jest to cyfrowy odpowiednik nut, z jakich korzystają muzycy.

```
<measure number="27" width="69.61">
  <harmony print-frame="no">
    <root>
      <root-step>F</root-step>
      <root-alter>1</root-alter>
      </root>
    <kind use-symbols="yes">diminished</kind>
    </harmony>
  <note>
    <rest/>
    <duration>24</duration>
    <voice>1</voice>
    </note>
  </measure>
```

*Rysunek 25 Przykład pliku MXML*

### 5.1.3 Graficzne

- Piano Roll – graficzna metoda reprezentacji muzyki, w której kolorowe paski reprezentują wysokość dźwięku (położenie w jednej z osi) oraz czas trwania (długość paska w drugiej z osi). Poniżej (zob. Rysunek 26) widać przykład zastosowania formatu piano roll do reprezentacji kolejnych nut w popularnym programie Synthesia. Kolejny obraz (zob. Rysunek 27) to tzw. piano roll z opisanymi osiami. Można wysnuć szybki wniosek, że w tym formacie często występują podłużne paski w osi czasu, reprezentują one długo trwające nuty. Brak z kolei tego typu elementów w osi częstotliwości, ponieważ byłyby one klastrami dźwięków, które brzmią specyficznie i nie stosuje się ich zbyt często w aranżacjach muzycznych.

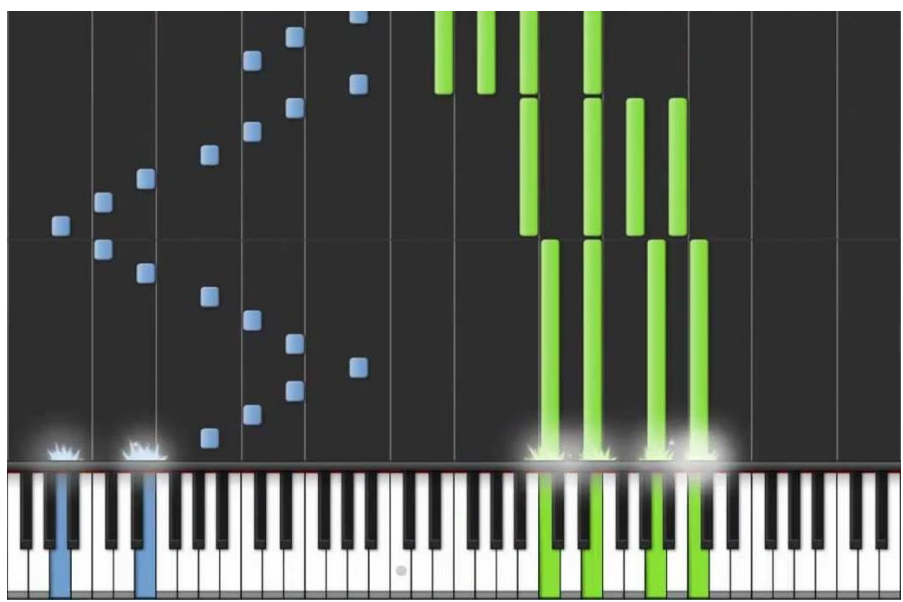

*Rysunek 26 Piano Roll w programie Synthesia [41]*

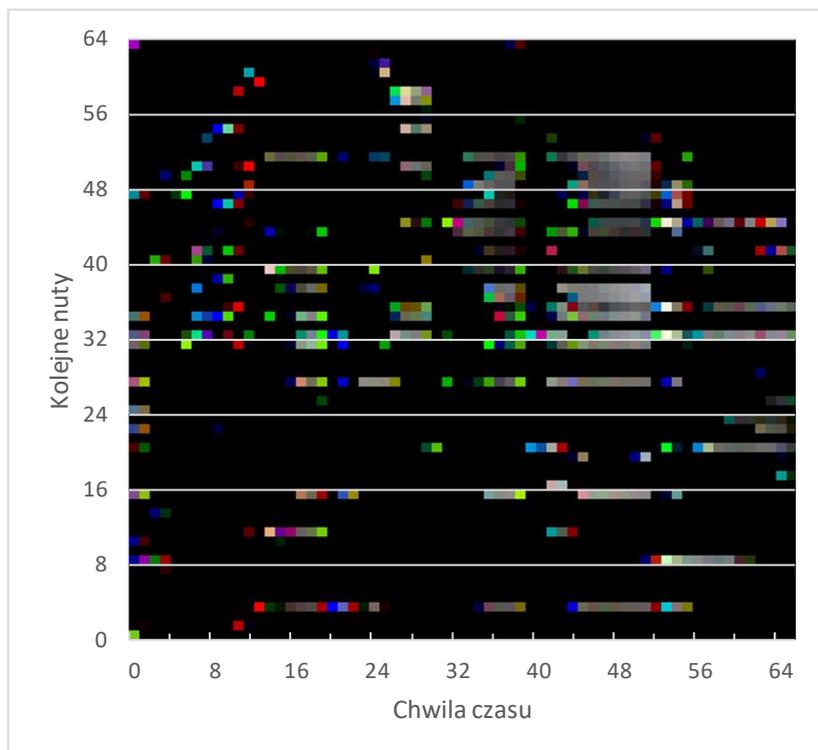

*Rysunek 27 Muzyka zapisana w formie obrazu piano roll*

- Spektrogram – wykres widma amplitudy składowych harmonicznych sygnału po czasie. To, czym wyróżnia ten sposób zapisu muzyki, jest to, że jako jedyny (poza WAV) reprezentuje całe spektrum częstotliwości w danej chwili czasu, co stanowi podstawę do analizy brzmienia dźwięku.

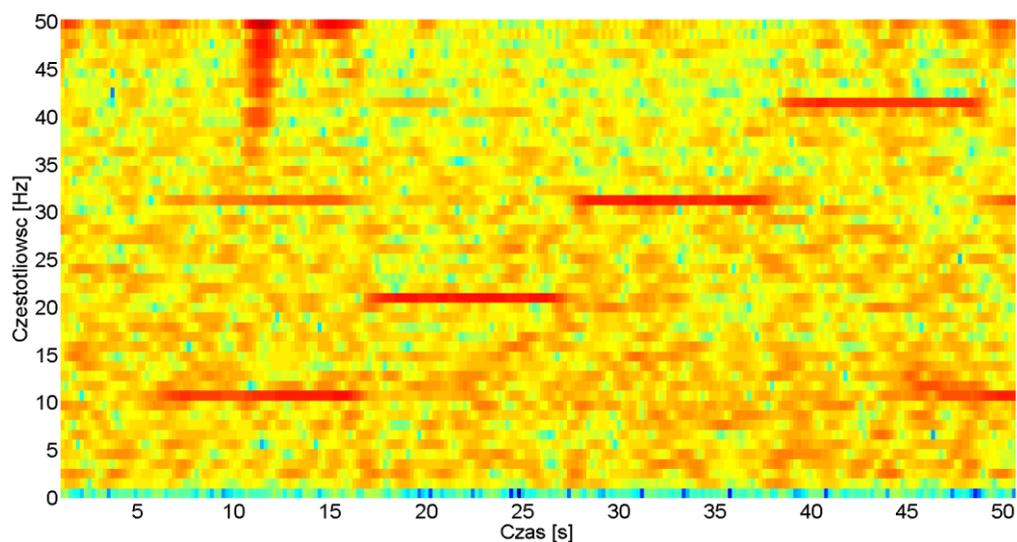

*Rysunek 28 Przykładowy spektrogram (Źródło [42]*

## 5.2 Muzyczne bazy danych

Muzyczne bazy danych, jakie możemy z naleźć w sieci to w głównej części zbiory MIDI, zbiory audio. Pliki audio można pobrać z serwisów takich jak YouTube. Pliki MIDI najczęściej można znaleźć na prywatnych repozytoriach. Poniżej wypisałem bazy plików MIDI, jakie znalazłem w sieci.

### 5.2.1 Kolekcja nagrań Dough-a McKenzie

Jest to zbiór [43] ponad 300 utworów trwających łącznie ponad 20 godzin nagrany przez wybitnego pianistę jazzowego. Charakteryzuje się dużą wirtuozerią, częstymi płynnymi zmianami tępa i dynamiki, bardzo bogatą i złożoną harmonią oraz rytmiką. W większości nagrań pianino gra solo, ale niektórych nagraniach występują też inne instrumenty takie jak kontrabas czy zestaw perkusyjny.

### 5.2.2 MAESTRO

Baza danych MAESTO [44] jest zbiorem ponad 172 godzin nagrań solowej, fortepianowej muzyki klasycznej w formie MIDI oraz WAV. Baza plików została stworzona na podstawie nagrań z międzynarodowego Piano-E-Competition, które od prawie 20 lat odbywają się w Minneapolis w USA [45] precyzja dopasowania między plikiem MIDI i odpowiadającym mu plikiem WAV waha się w okolicach 3ms, czyli na granicy ludzkiej percepcji. Pliki MIDI, jakie zawiera baza są bardzo dobrej jakości, brak tam zbędnych zdarzeń. Muzyka, jaką reprezentuje baza danych, jest bardzo wirtuozerska, jej struktura harmoniczna jest poukładana, zgodna z zasadami muzyki klasycznej – prostsza niż w przypadku muzyki jazzowej. Również struktura rytmiczna utworów jest bardziej schematyczna niż w muzyce jazzowej. Dzięki takiej charakterystyce muzyki łatwiej jest wykryć powtarzające się struktury oraz zależności między nimi. Nieprzetworzone pliki MIDI po uruchomieniu w programie DAW (digital audio workstation) Ableton z wykorzystaniem odpowiedniej wtyczki brzmiały niemal tak jak prawdziwe nagranie.

### 5.2.3 Pozostałe

Przeglądając Internet w poszukiwaniu plików MIDI z nagraniami jazzowych utworów, natrafiłem na dziesiątki stron zawierających od kilkunastu do kilkuset takich plików. Ich wyszukiwanie pobieranie i sortowanie to dość żmudne zajęcie. Analiza tak zebranych plików pozwala wysnuć kilka wniosków. Przede wszystkim jest to zbiór bardzo różnorodny pod względem stylów, jakości wykonania, instrumentarium i jakości plików - chodzi tu o ilość zbędnych zdarzeń.

## 5.3 Zastosowane rozwiązania

W moim projekcie zdecydowałem się użyć formatu graficznego Piano Roll, ponieważ najlepiej reprezentują on strukturę rytmiczno-czasową i harmoniczno-melodyczną. W związku z tym, że tego typu dane występują w Internecie w bardzo małej liczbie, zdecydowałem się uzyskać je poprzez przekształcenie plików MIDI, których popularność jest znacznie większa. Zdecydowałem się skorzystać z bazy danych MAESTRO, zawiera nieco prostsze utwory pod względem harmonicznym i rytmicznym niż zbiór Dough'a McKenzie, ale jest 8 razy większa oraz zawiera lepszej jakości pliki MIDI – nie posiada dodatkowych instrumentów i zbędnych zdarzeń.

# 6 Generowanie muzyki z wykorzystaniem DCGAN

Głównym założeniem projektu jest stworzenie programu zdolnego do generowania muzyki, która posiada złożoną struktur czasową – rytm oraz dźwiękową – melodia i harmonia, przez co należy reprezentować ją w postaci dwuwymiarowej, aby oddać jej wewnętrzną strukturę, innym przykładem struktur 2D są obrazy. Przekształcenie muzyki na obraz stanowi dużą część projektu, jest niej poświęcony poniższy rozdział. Typowym podejściem w przetwarzaniu obrazów jest zastosowanie głębokich sieci splotowych. Do generowania obrazów stosowane są najczęściej generatywne sieci przeciwstawne – GAN. W moim projekcie zdecydowałem się wykorzystać hybrydę tych dwóch rozwiązań, czyli głębokie splotowe generatywne sieci przeciwstawne (zob. Rysunek 29) (ang. deep convolutional generative adversarial networks - DCGAN)

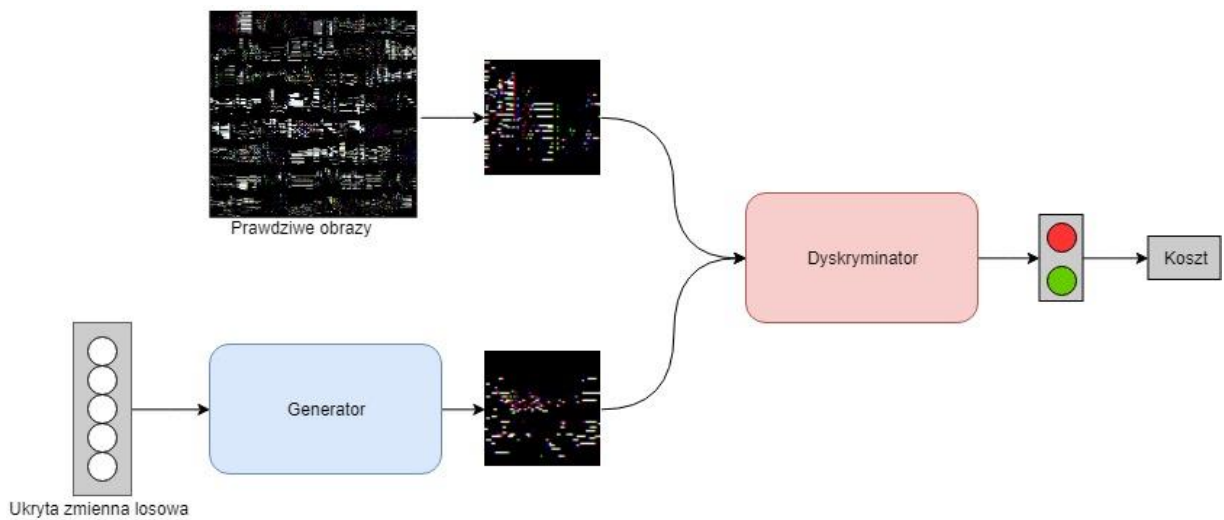

*Rysunek 29 Struktura sieci DCGAN*

## 6.1 Założenia projektowe – szczegóły techniczne

Założenia, jakie sobie wyznaczyłem, to wydobycie jedynie niezbędnych danych, to znaczy partii fortepianu, następnie znormalizowania tempa i skwantyzowania wartości rytmicznych, a na koniec złączenie w jeden duży plik, z którego generowane były obrazy, które utworzą zbiór treningowy. Przyjęcie takiej koncepcji było spowodowane ograniczonym rozmiarem obrazów wejściowych oraz potrzebą uzyskania jak najbardziej wyrazistych obrazów w celu ułatwienia ekstrakcji cech i przyspieszenia nauki. Niesie to ze sobą szereg konsekwencji. Przede wszystkim ograniczenie dynamiki utworu i wyprostowanie rytmu powoduje, że utwór brzmi trochę „sztywno", traci w pewnym stopniu swoją płynność i w odbiorze jest lekko syntetyczny. Nie wpływa to jednak na samą treść muzyczną, jaką utwór niesie. Nadal słychać spójne frazy, puls utworu i strukturę harmoniczną. Utwory, na których chciałem trenować sieć, dobierałem patrząc na takie cechy, jak złożona harmonia i rytm, ponieważ chciałem zbadać, jak można odzwierciedlić strukturę muzyczną w pełnej krasie, a nie jedynie jej najbardziej podstawową postać, z jaką często mamy do czynienia w muzyce popularnej. Dlatego mój wybór padł na jazz i muzykę klasyczną. Oczekiwany wynik działania sieci to generowanie danych podobnych do przetworzonych danych treningowych, co oznacza, że one również będą brzmiały nieco syntetycznie. To, czego należy w nich szukać to złożona i ciekawa struktura harmoniczna, melodyczne frazy, niebanalny rytm i wyczuwalny puls.

## 6.2 Stos technologiczny

W tym rozdziale pokrótce opiszę wybrane elementy stosu technologicznego. Wybrałem środowisko programistyczne PyCharm, ponieważ oprócz podstawowych cech takiego środowiska posiada wbudowaną integrację z systemem kontroli wersji oraz potrafi zarządzać wirtualnymi środowiskami. Skorzystałem z wersji profesjonalnej dzięki studenckiej subskrypcji. Język programowania zastosowany w projekcie to python w wersji trzeciej. Zawiera szereg bibliotek przydatnych zarówno w przetwarzaniu danych, jak i w implementacji sieci neuronowych. Główna biblioteka do tworzenia sieci neuronowej w moim projekcie to PyTorch, intensywnie rozwijana, pozwalająca na integrację w wewnętrzną strukturę modelu, bardzo elastyczna i dość dobrze udokumentowana. Manager środowisk, z jakiego skorzystałem co Miniconda. Pozwala na tworzenie zupełnie niezależnych od siebie środowisk oraz dużą swobodę w manipulacji zainstalowanymi pakietami. Do archiwizacji i kontroli wersji posłużył mi Git oraz zdalne repozytorium kodu GitHub również ze studencką licencją pozwalającą tworzyć prywatne repozytoria.

## 6.3 Wstępne przetwarzanie danych

Proces wstępnego przetworzenia danych rozbiłem na kilka kroków przedstawionych w poniższym diagramie (zob. Rysunek 30). Każdy z tych procesów został zaimplementowany w postaci skryptu w języku Python z wykorzystaniem biblioteki Numpy do złożonych obliczeń macierzowych.

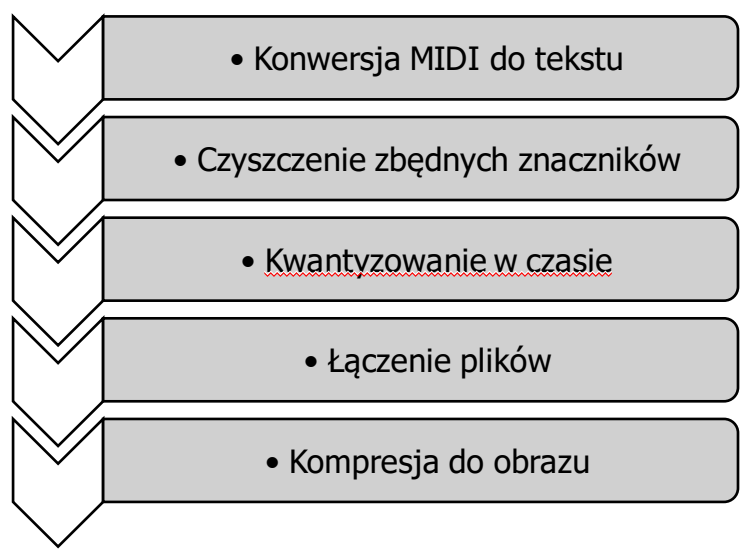

*Rysunek 30 Algorytm przetwarzania danych*

### 6.3.1 Konwersja MIDI do formy tekstowej

Plik MIDI ma formę protokołu, w którym kolejne linie to zdarzenia reprezentujące akcje klawiszy lub innych kontrolek. Dodatkowo jest on przechowywany w formie binarnej, więc niezbędna była konwersja do formy tekstowej.

```
MFile 1 2 384
MTrk
0 Tempo 500000
0 TimeSig 4/4 24 8
1 Meta TrkEnd
TrkEnd
MTrk
0 PrCh ch=1 p=0
806 On ch=1 n=42 v=71
909 Par ch=1 c=64 v=37
924 Par ch=1 c=64 v=50
939 Par ch=1 c=64 v=58
955 Par ch=1 c=64 v=62
970 Par ch=1 c=64 v=65
986 Par ch=1 c=64 v=68
. . .
```

### 6.3.2 Czyszczenie pliku

Pozbycie się niepotrzebnych danych (informacje o brzmieniach, komentarze, itp.), ograniczenie dynamiki dźwięków do wartości maksymalnej, skwantyzowanie do 30 ms ( wartość szesnastki w tempie 120 bpm).

```
MFile 1 1 120
MTrk
0 On ch=1 n=49 v=127
0 On ch=1 n=77 v=127
30 On ch=1 n=77 v=0
30 On ch=1 n=73 v=127
30 On ch=1 n=73 v=0
30 Par ch=1 n=64 v=127
60 On ch=1 n=68 v=127
60 On ch=1 n=49 v=0
60 On ch=1 n=68 v=0
90 On ch=1 n=73 v=127
. . .
```

### 6.3.3 Łączenie

Połączenie wszystkich plików do jednego, dużego, usunięcie niepotrzebnych pauz, kompresowanie do obrazów 64x64 (zob. Rysunek 31). Pozioma oś obrazu reprezentuje czas, a pionowa kolejne nuty. Skala fortepianu (88 klawiszy) została ograniczona do 64 klawiszy (wysokość obrazu) poprzez transponowanie o oktawę, nut spoza zakresu. Kompresji poddane zostały poddane też długości trwania nut. Najkrótsza nuta możliwa do uzyskania po kwantyzacji, czyli trwająca 30 ms w tempie 120 bpm szesnastka, zakodowana jest nie na jednym pikselu, lecz na jednym z trzech kanałów, odpowiadających danemu pikselowi. Przestrzeń czasowa została trzykrotnie rozciągnięta i każda z nut została zakodowana z wykorzystaniem wszystkich trzech subpikseli bitmapy, dlatego efektem jest kolorowy obraz, składający się z 8 kolorów (R – czerwony G – zielony B – niebieski C – cyjan M – magenta Y – żółty K – czarny – W – biały). Jeżeli nuta trwała dwie ósemki i zaczęła się na początku utworu, to jej zapis składał się z ustawienia maksymalnej wartości dla piksela R i G, co w wyniku daje kolor żółty – Y. Jeżeli nuta trwała dłużej niż trzy szesnastki i zapełniła 3 subpiksele należące do jednego piksela to ten piksel zaświecił się na biało.

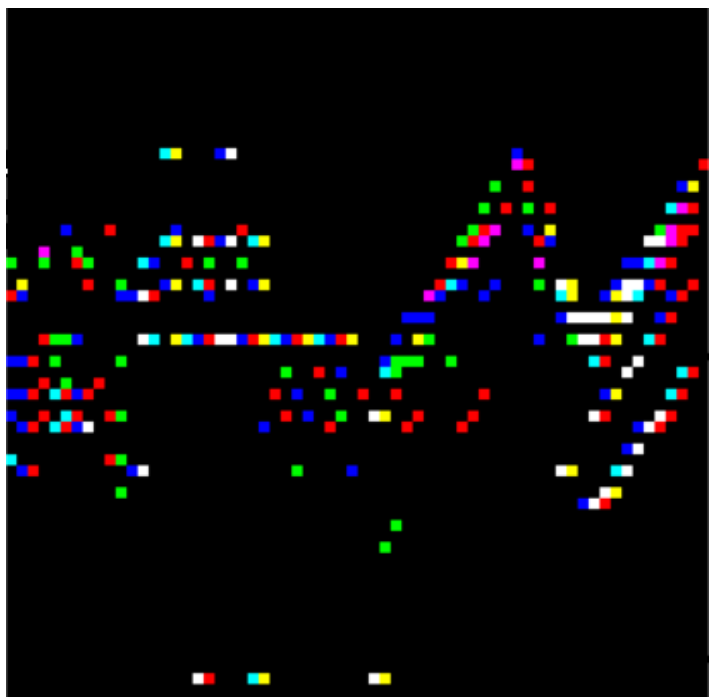

*Rysunek 31 Przykład zakaodwanego na kanałach RGB pliku muzycznego MIDI*

### 6.3.4    Błędy w plikach

Część plików posiadała błędy w postaci niezakończonych zdarzeń kliknięcia pedała (zob. Rysunek 32), co powodowało, że na ekranie pojawiały się bardzo długie fragmenty białych pasków, które reprezentowały kolejno wciskane nuty, które zostały przedłużone przez zbyt długo wciśnięty pedał sustain (ostatni z trzech pedałów w fortepianie służący do wydłużania czasu trwania granych dźwięków). Zamiast usuwać wygenerowane obrazy, w których wystąpił błąd postanowiłem zmniejszyć skutki tego typu błędów poprzez automatyczne wyłączanie działania pedału sustain po 3 sekundach po jego włączeniu. Czas ten wybrałem, ponieważ pianiści używają go w większości przypadków przez krótszą chwilę, a ewentualnie skrócenie wybrzmienia końcowego akordu nie wpływa na strukturę harmoniczną i rytmiczną utworu. Czas ten jest natomiast na tyle krótki, że wystąpienie błędu nie jest bardzo odczuwalne, ponieważ jego wpływ został znacząco skrócony. Pozwolił to wygenerować obrazy, które reprezentują muzykę zgodnie z moimi założeniami (zob. Rysunek 33).

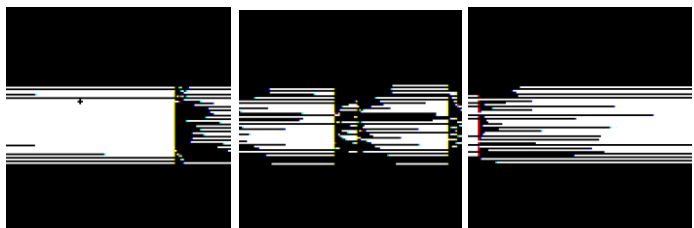

*Rysunek 32 Przykład danych treningowych zawierających opisany błąd*

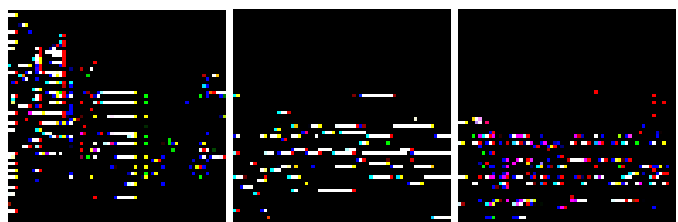

*Rysunek 33 Dane treningowe*

## 6.4 Charakterystyka Danych Treningowych

Tak powstałe obrazy można przekonwertować bezstratnie na plik MIDI z wykorzystaniem stworzonego przeze mnie skryptu. Muzyka z tak powstałego pliku MIDI ma nieco poszarpany i nierównomierny rytm (wynik kwantyzacji) w porównaniu do nieprzetworzonych plików MIDI oraz dość toporną dynamikę (w związku z przyjęciem binarnej wartości dynamika każdego z dźwięków). Fragment muzyczny odpowiadający każdemu z obrazków trwa niewiele ponad 20 sekund. Przetworzenie całej bazy danych daje w efekcie ponad 25 tysiące obrazów. Opracowałem też algorytm dekompresji z bitmapy do pliku tekstowego, a następnie do MIDI, dzięki czemu będzie można odsłuchać wynik działania sieci neuronowej.

## 6.5 Uczenie się sieci

Sieć neutronowa jest zaimplementowana w bibliotece PyTorch, a do obliczeń skorzystałem z chmury obliczeniowej Google Colaboratory. Postanowiłem wykorzystać do generowania muzyki implementację [46] sieci Deep Convolutional Generative Adversarial Network (DCGAN) opisaną w dokumencie [5]. Dzięki zastosowaniu biblioteki CUDA firmy NVidia do obliczeń na GPU uczenie trwało kilka razy krócej, niż by to miało miejsce z wykorzystaniem CPU. Karta graficzna, która wykonywała obliczenia na serwerze to NVidia Tesla K80 – jedna z topowych kart graficznych dostępnych na rynku. Dostępne było też 12GB pamięci RAM karty graficznej. Podsumowując: proces nauki trwał 50 tys. iteracji, co przełożyło się na około 11h pracy maszyny.

```
Gen RAM Free: 12.8 GB  | Proc size: 141.7 MB
GPU RAM Free: 11441MB | Used: 0MB | Util   0% | Total 11441MB
```

Generator jest siecią splotową, składającą się z na przemian występujących warstw splotowej, normalizacyjnej i aktywacji (zob. Rysunek 34).

```
Generator(
  Sequential(
    (0): ConvTranspose2d(100, 512)
    (1): BatchNorm2d(512)
    (2): ReLU()
    (3): ConvTranspose2d(512, 256)
    (4): BatchNorm2d(256)
    (5): ReLU()
    (6): ConvTranspose2d(256, 128)
    (7): BatchNorm2d(128)
    (8): ReLU()
    (9): ConvTranspose2d(128, 64)
    (10): BatchNorm2d(64)
    (11): ReLU()
    (12): ConvTranspose2d(64, 3)
    (13): Tanh()
  )
)
```

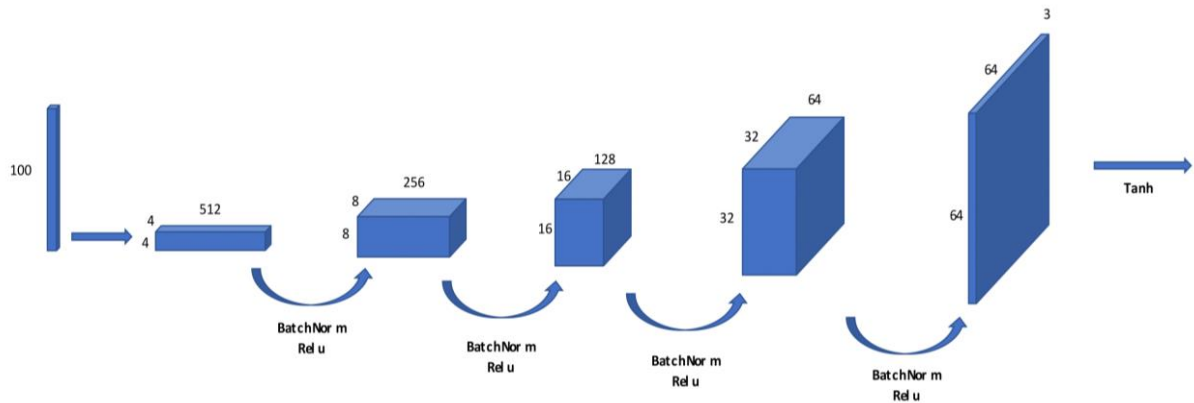

*Rysunek 34 Struktura generatora*

Dyskryminator (zob. Rysunek 35) ma podobną strukturę do generatora, lecz kolejne warstwy zmniejszają swój wymiar wejściowy, względem warstwy poprzedniej, by na końcu otrzymać binarną wartość decydującą o tym, czy obraz wejściowy był generowany przez generator.

```
Discriminator(
  Sequential(
    (0): Conv2d(3, 64)
    (1): LeakyReLU()
    (2): Conv2d(64, 128)
    (3): BatchNorm2d(128)
    (4): LeakyReLU()
    (5): Conv2d(128, 256)
    (6): BatchNorm2d(256)
    (7): LeakyReLU()
    (8): Conv2d(256, 512)
    (9): BatchNorm2d(512)
    (10): LeakyReLU()
    (11): Conv2d(512, 1)
    (12): Sigmoid()
  )
)
```

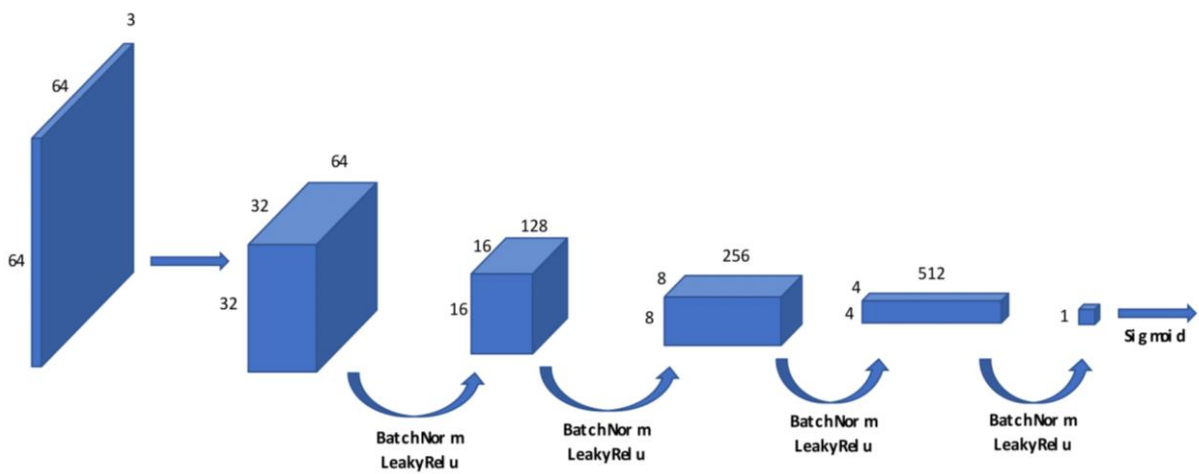

*Rysunek 35 Struktura dyskryminatora*

Zgodnie z zaleceniami w [5] warstwa splotowa i normalizująca są inicjalizowane wartościami z rozkładu normalnego o odpowiednich parametrach podanych, jako argumenty w poniższym kodzie.

```
def weights_init(m):
    classname = m.__class__.__name__
    if classname.find('Conv') != -1:
        nn.init.normal_(m.weight.data, 0.0, 0.02)
    elif classname.find('BatchNorm') != -1:
        nn.init.normal_(m.weight.data, 1.0, 0.02)
        nn.init.constant_(m.bias.data, 0)
```

Wykres funkcji kosztu prezentuje poniższy obraz (zob. Rysunek 36). Funkcja kosztu w pierwszych 5 tys. iteracjach wskazuje na to, że doszło do odwrócenia procesu nauczania. Generator trafił do lokalnego minimum, z którego nie potrafił się wydostać, do momentu, w którym jego struktura znacznie się zmieniła. W dalszych iteracjach funkcja kosztu generatora spada, a następnie utrzymuje się w stałym przedziale, natomiast funkcja kosztu dyskryminatora przez większość czasu ma wartość równą lub bliską zeru. Oznacza to również, że GAN zatrzymał się w jednym z lokalnych minimów, z którego już nie potrafił się wydostać, pomimo tego wyniki okazały się na tyle dobre, że były nieodróżnialne na pierwszy rzut oka.

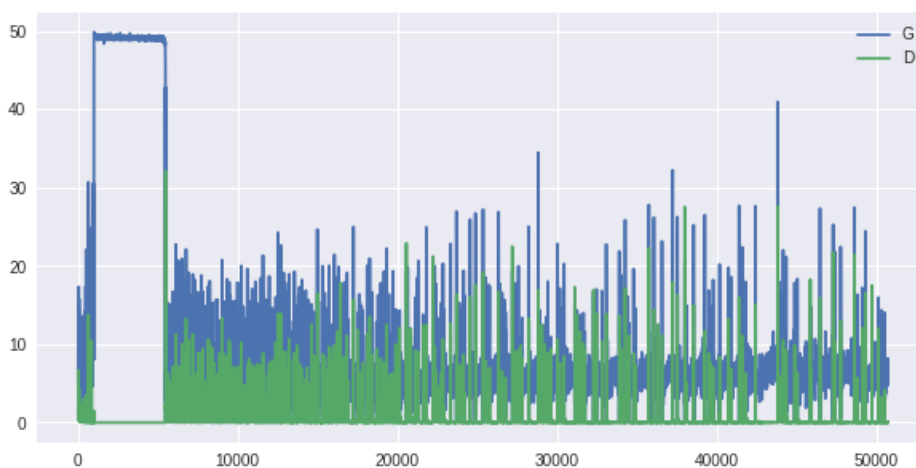

*Rysunek 36 Wykres funkcji kosztu generatora i dyskryminatora*

## 6.6 Pierwsze uzyskane wyniki

Efekt uzyskany poprzez przetworzenie całej bazy danych (25 tys. obrazów) w 250 epokach zamieszczony jest niżej (zob. Rysunek 37). Oba rysunki prezentują tablice 64 obrazów prawdziwych i 64 obrazów wygenerowanych przez sieć neuronową.

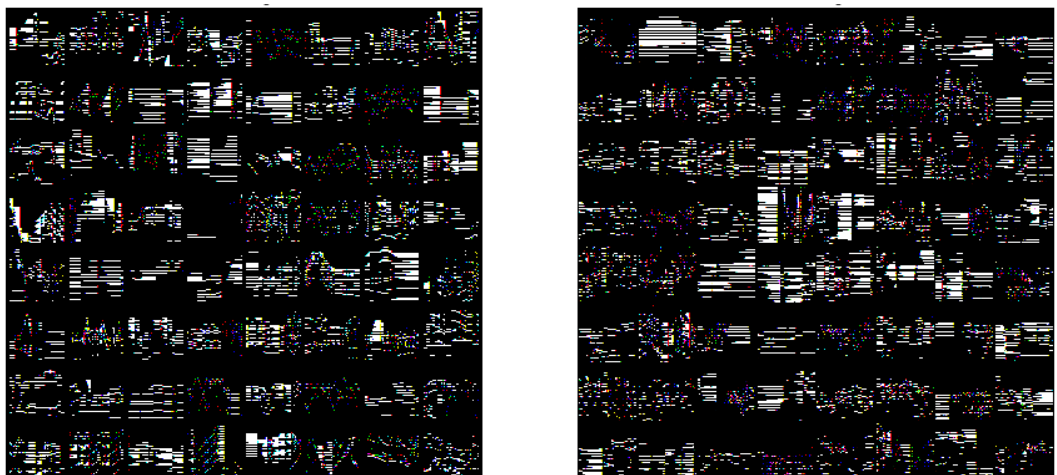

*Rysunek 37 Obrazy z bazy treningowej (po lewej) obrazy wygenerowane (po prawej)*

Powyższe porównanie pokazuje, że na pierwszy rzut oka nie jest możliwe odróżnienie obrazów rzeczywistych od wygenerowanych. Posiadają one zbliżone zestawy kolorów oraz podobne zagęszczenie i kształt wzorów. Poniżej zamieszczam przykładowy obraz wygenerowany przez sieć.

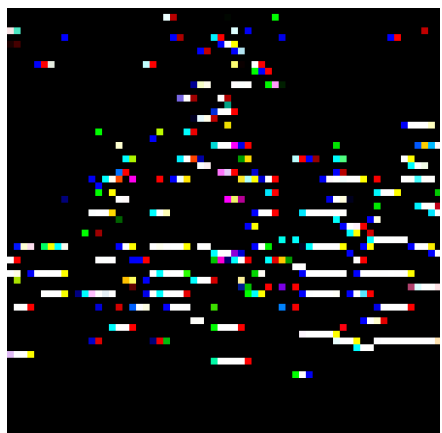

*Rysunek 38 Przykładowy obraz wygenerowany przez sieć*

Tak powstałe obrazy przetworzyłem przygotowanym przeze mnie skryptem służącym do dekompresji do formatu tekstowego, a następnie do pliku MIDI. Po odsłuchaniu plików MIDI stworzonych przez generator w ostatniej epoce wyciągam następujące wnioski:

- W tak otrzymanej muzyce występują stałe i powtarzające się struktury rytmiczne, co jest charakterystyczne dla muzyki klasycznej, jaką „karmiona" była sieć.
- Struktura harmoniczna ma podobną złożoność do danych treningowych. Zaskakująco często pojawiają się popularne struktury harmoniczne (akordy molowe, durowe, akordy kwartowe), a także oddzielna linia basowa, rozłożone przebiegi – arpeggia.
- Nieco rzadziej występują dobrze brzmiące kadencje. Kilka przykładów zawierały rozwiązania dominant lub akord toniczny na koniec frazy co było dla mnie ogromnym zaskoczeniem, ponieważ jest to trudny do wykrycia przez swą złożoność, ale bardzo istotny element kompozycji muzycznej.

## 6.7 Dodatkowe eksperymenty

Postanowiłem przeprowadzić kilka dodatkowych eksperymentów, aby sprawdzić, w jakim stopniu mogę poprawić działanie sieci oraz jakie jej elementy mają największy wpływ na jakość generowanych obrazów, a dokładniej muzyki, jaka z nich jest uzyskiwana.

### 6.7.1   Dodatkowe iteracje

Efekty uzyskane po wykonaniu dodatkowych 20 tys. iteracjach nie wprowadziły znacznej poprawy. Funkcja kosztu generatora nadal oscyluje wokół pewnej ustalonej wartości (zob. Rysunek 40), a dyskryminator prawie bezbłędnie odróżnia falsyfikaty od oryginałów, o czym świadczy jego bliska zeru funkcja kosztu. Muzyka generowana po dodatkowych iteracjach (zob. Rysunek 41) brzmi podobnie i nadal jest chaotyczna. Aby wyraźnie poprawić jakość generowanej muzyki, należy zaingerować w bazę danych, strukturę sieci lub jej parametry uczenia.

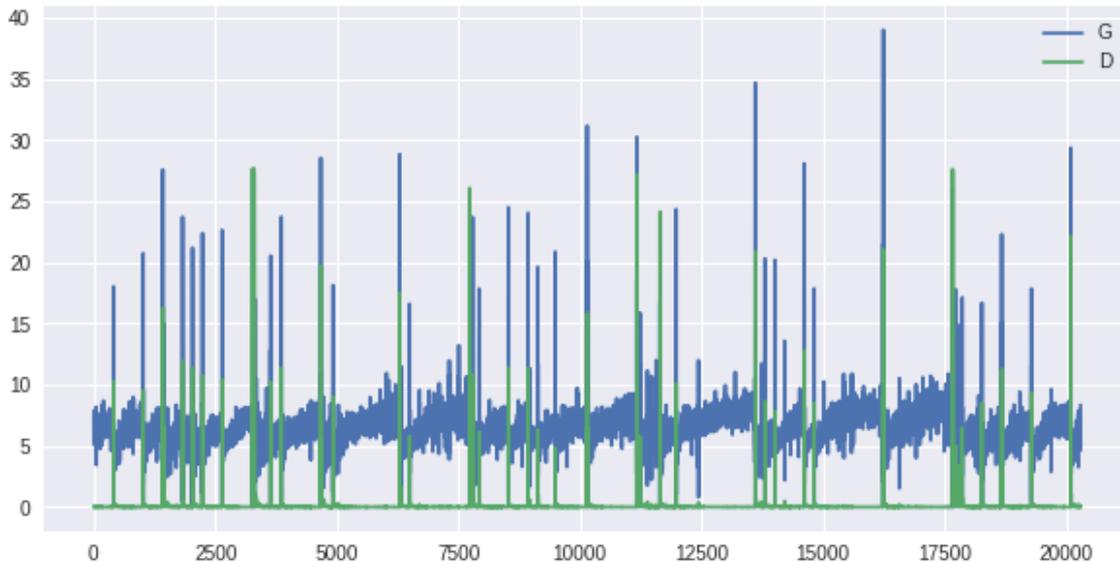

*Rysunek 39 Funkcja kosztu w dodatkowych 20 tys. iteracjach*

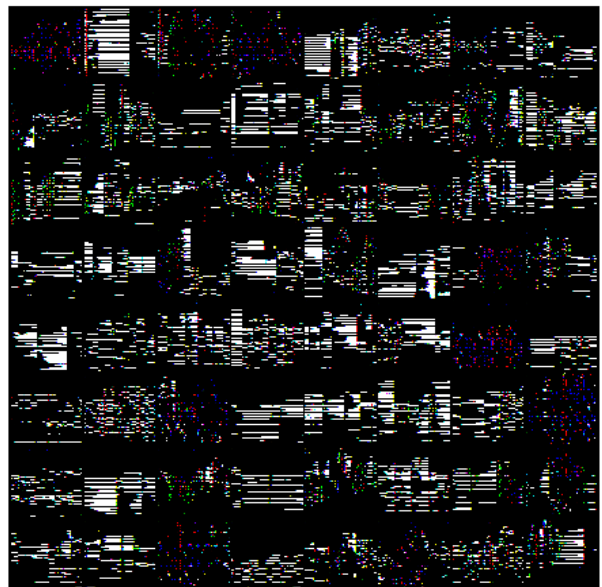

*Rysunek 40 Obrazy wygenerowane w ostatniej iteracji.*

### 6.7.2 Inna baza danych

W tym eksperymencie zamiast bazy obrazów uzyskanej z bazy plików MIDI - MAESTRO [44] (~25 tys. obrazów) użyłem bazy (~2,5 tys. obrazów), którą stworzyłem na podstawie zbioru plików MIDI Dough-a McKenzie [43] – jest to zbiór utworów jazzowych. Baza miała mniejszy rozmiar, dzięki czemu w podobnym czasie mogłem wykonać znacznie większą liczbę iteracji. W funkcji kosztu po około 120 tys. iteracjach (zob. Rysunek 42) doszło do zjawiska odwrócenia się procesu uczenia i sieć ugrzęzła w bardzo słabym minimum lokalnym, czego efekty objawiły się generowaniem obrazów, które po przetworzeniu na muzykę okazały się bezsensowną i gęstą plątaniną dźwięków (zob. Rysunek 42).

Obrazy generowane przez sieć tuż przed znacznym pogorszeniem jej sprawności były bardzo podobne do tych uzyskanych w pierwotnym eksperymencie, co jedynie potwierdza, że zmiana źródła danych na takie z innym gatunkiem muzycznym nie zmienia jakości czy charakteru generowanej muzyki. Widoczna poprawa mogłaby być widoczna, gdybyśmy mieli do czynienia z bardzo ubogimi harmonicznie i rytmicznie fragmentami muzycznymi, lecz jest to rozbieżne z założeniami tego projektu. Warte uwagi jest też to, że wykorzystana w tym eksperymencie baza była około 10 razy mniejsza niż w pierwotnym eksperymencie, co miało wpływ na osiągnięcie mniejszej ogólności przez generator.

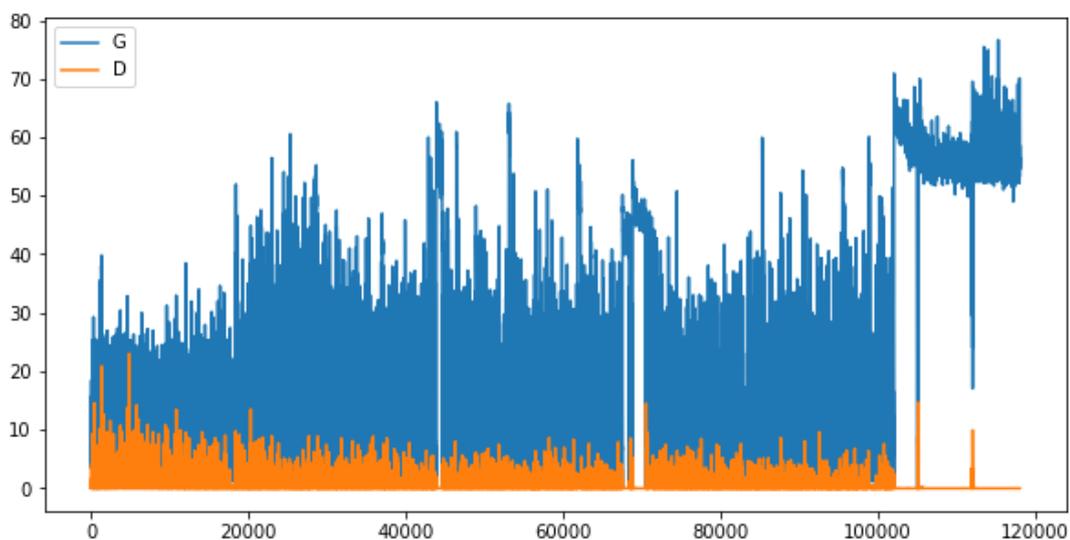

*Rysunek 41 Funkcja kosztu przy uczeniu na mniejszej bazie*

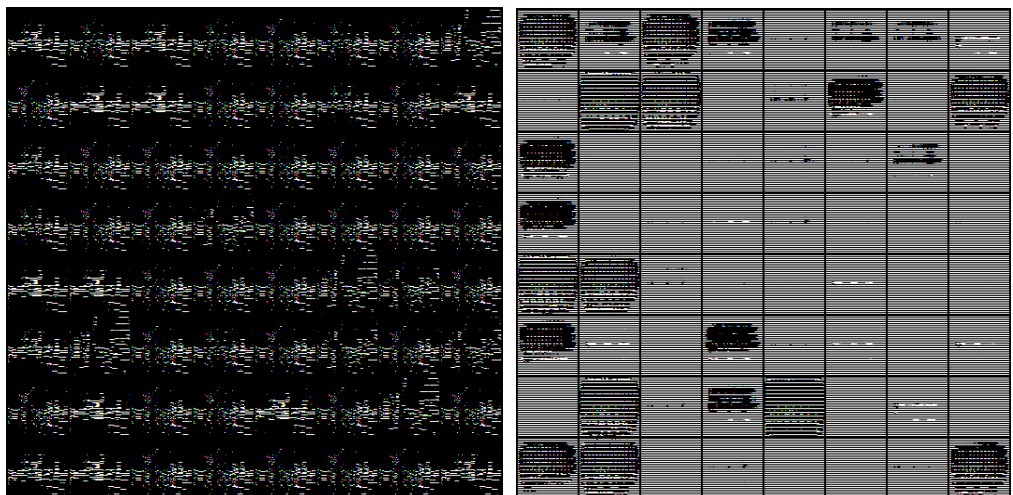

*Rysunek 42 Obrazy generowane przed (po lewej) i po (po prawej) odwróceniu procesu uczenia.*

### 6.7.3 Pełen zakres dynamiczny.

Zgodnie z założeniami projektu, dynamika utworów była kwantyzowana do wartości maksymalnej. Utwory po procesie wstępnej obróbki miały płaską dynamikę ustawioną na jej maksymalną wartość, przez co straciły w pewnym stopniu swój estetyczny walor. Baza, jakiej użyłem to ta z eksperymentu bazowego (25tys. obrazów reprezentujących muzykę klasyczną). W tym eksperymencie porzuciłem to założenie na rzecz weryfikacji tego, w jakim stopniu szerokość zakresu dynamicznego może wpłynąć na proces uczenia. Wart nadmienienia jest fakt, że obrazy generowane w bazowym eksperymencie posiadały około 1% pikseli, które nie miały pełnego natężenia, co odpowiadało nutom o pośredniej wartości dynamiki. Możliwe, że to właśnie po tej cesze dyskryminator oceniał prawdziwość danych.

Wyniki eksperymentu dały wynik, jakiego do tej pory jeszcze nie było, mianowicie koszt dyskryminatora po pewnym czasie ustalił się na pewnej dużej wartości, a koszt generatora spadł do zera (zob. Rysunek 44). Oznacza to, że generator nauczył się tworzyć falsyfikaty, które dla dyskryminatora były nie do odróżnienia od oryginałów. Na pierwszy rzut oka, również nie można odróżnić obrazów generowanych od oryginalnych (zob. Rysunek 45), co stanowi o pewnym postępie w rozwoju projektu oraz zachęca do refleksji nad początkowymi założeniami. Niestety takie zachowanie się sieci, z pozoru dobrze się zapowiadające, oznacza, że generator nie rozwijał się dalej, ponieważ znalazł prosty sposób, żeby oszukać dyskryminator. Efekty takiej pracy generatora można ocenić, słuchając muzyki, którą wytworzył. Jest ona jeszcze bardziej chaotyczna niż w eksperymencie początkowym, zdecydowanie trudniej odnaleźć w niej takie elementy jak struktury harmoniczne, czy płynne lub melodyczne frazy.

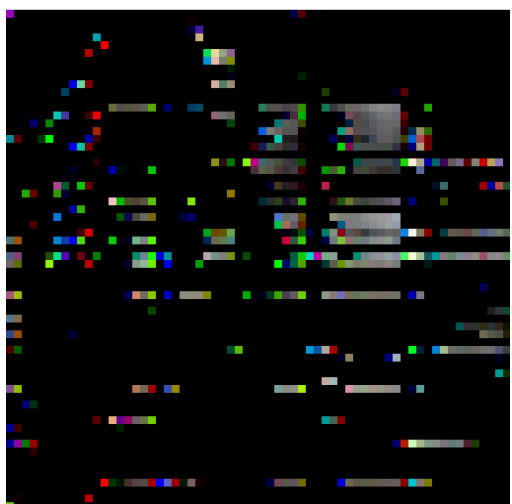

*Rysunek 43 Obraz treningowy - pełen zakres dynamiki*

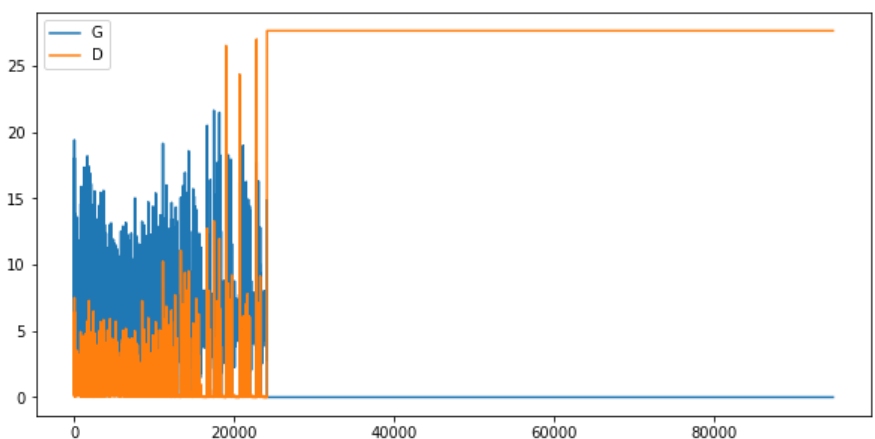

*Rysunek 44 Funkcja kosztu w eksperymencie z pełnym zakresem dynamiki*

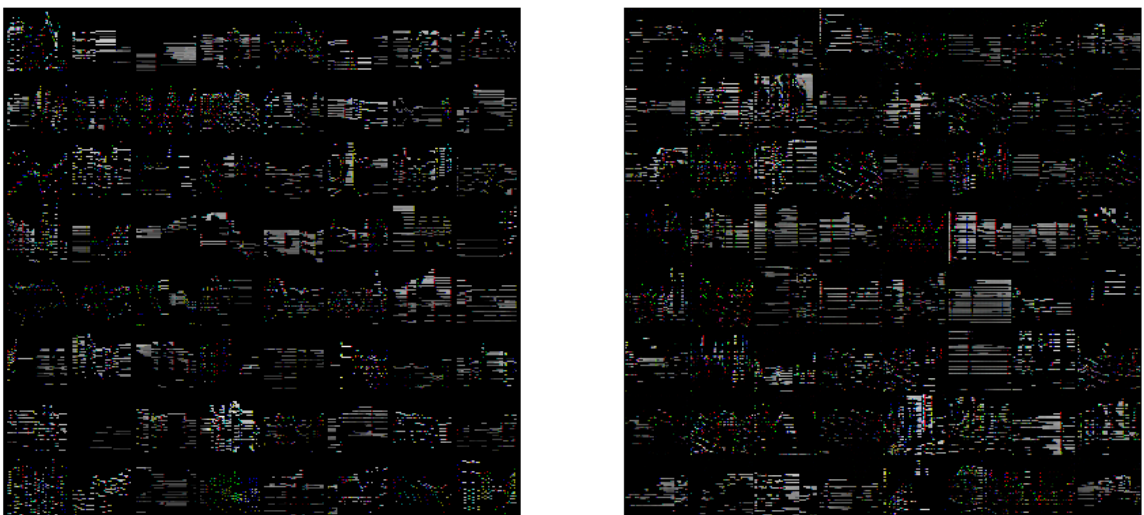

*Rysunek 45 Prawdziwe obrazy (po lewej), fałszywe (po prawej) - pełny zakres dynamiki*

## 6.8 Spełnienie wymagań projektowych

Założenia projektowe zostały spełnione. Program generuje frazy, w których słyszalne są fragmenty zawierające rozbudowane akordy (akordy durowe, mollowe, dominanty, w eksperymencie na utworach jazzowych parę razy pojawiły się typowe dla bluesa jazzowego złożone akordy alterowane kończące frazę). Rytm mimo, że chaotyczny w każdym z utworów był zbliżony na przestrzeni trwania danego fragmentu. Występowanie linii górnego głosu, która koresponduje z wewnętrzną strukturą harmonii, może świadczyć o świadomości sieci o istnieniu melodii i jej zależnością z resztą utworu.

# 7 Podsumowanie

## 7.1 Porównanie z podobnymi projektami

Projekty związane z generowaniem muzyki najczęściej bazują na jakichś danych wejściowych i od nich uzależniają swój wynik działania. Na przykład poprzez stworzenie melodii do konkretnych akordów lub wygenerowanie melodii zbliżonej do danej. W moim projekcie wykorzystałem jedynie losowy szum na wejściu generatora do stworzenia unikalnej próbki. Efekt działania mojego algorytmu to 20-sekundowy fragment muzyczny. W większości projektów generowany fragment trwa o wiele krócej. Złożoność harmoniczna utworów, które generuje mój algorytm jest o wiele większa niż w wielu projektach Magenty. Chaotyczność rytmu i melodii to cechy, które w największym stopniu wpływają na mniejsze zrozumienie odsłuchiwanego materiału, który jest wynikiem działania generatora mojej sieci. Główne wysiłki tej pracy były kierowane w zbadanie wpływu różnych parametrów algorytmu na jakość wyników niż na samo zwiększenie jakości. Osiągnięcie przewagi na gruncie złożoności harmonicznej i długości próbki jest wynikiem zastosowania bazy danych wirtuozerskich utworów oraz przekształcenie jej na obrazy, które mogą przechować więcej informacji niż pliki tekstowe.

## 7.2 Napotkane trudności

Dużą trudnością były kwestie sprzętowe. Architektura DCGAN jest złożona, operuje na dużej ilości danych, więc wymagała przeprowadzenia obliczeń z wykorzystaniem GPU i architektury obliczeniowej CUDA. Karta graficzna, jaką posiadałem w laptopie - NVidia 740M, nie była zdolna do przeprowadzania tak złożonych obliczeń. W związku z tym byłem zmuszony do skorzystania z chmury obliczeniowej Google Colaboratory. Maszyna jaką zapewniała ta usługa chmurowa posiadała kartę graficzną NVidia Tesla K80, czyli urządzenie o potężnej mocy obliczeniowej, dzięki temu udało się przeprowadzić wystarczającą liczbę iteracji w ograniczonym przez Google czasie 12 h nieprzerwanego działania maszyny wirtualnej. Dalsze eksperymenty wymagały nieco dłuższej pracy systemu, więc potrzebowałem dostępu do lokalnej maszyny z mocną kartą graficzną. Dostałem dostęp do maszyny w instytucie dyplomującym z mocną kartą NVidia 780Ti, lecz wystąpił problem w momencie przenoszenia danych treningowych na pamięć karty. Okazało się, że w tym momencie pobór mocy z karty przekraczał możliwości zasilacza. Dwa rozwiązania, jakie mogłem zastosować w tej sytuacji, to wymiana zasilacza lub ograniczenie mocy karty graficznej, niestety karty graficzne NVidia serii 7 nie posiadają sterowników, które na to pozwalają, dlatego zdecydowałem się na zakup prywatnego laptopa z mocniejszą kartą graficzną – NVidia 1050Ti, co pozwoliło na dokończenie eksperymentów.

## 7.3 Możliwość rozwoju

Aby poprawić jakość generowanej muzyki warto byłoby przetestować inne architektury sieci, takie jak CNN, RNN, LSTM, VAE. Z pewnością tego typu badania będą wymagały ogromnych zasobów obliczeniowych i czasowych, w związku z tym najlepiej przenieść obliczenia na lokalny serwer lub skorzystać z innych rozwiązań chmurowych. Dodatkowo tak jak w każdym projekcie związanym z uczeniem maszynowym, warto poszerzyć zbiór uczący, poprawić jego jakość oraz zastanowić się nad ewentualną augmentacją.

## 7.4 Co zrobiłbym inaczej

Z pewnością nie traciłbym czasu na kolekcjonowanie własnej bazy danych, ale skorzystałbym z rozwiązania jakim jest gotowa baza danych w moim przypadku baza MAESTRO. Zastanowiłbym się również nad skorzystaniem z innej biblioteki niż Pytorch, na przykład Keras. Wysokopoziomowe API jakie zapewnia Keras to duża zaleta dla osób zaczynających pracę z uczeniem maszynowym, Pytorch pozwala na swobodną ingerencję w strukturę sieci i obliczenia na niej wykonywane, lecz odbywa się to kosztem niższej abstrakcji, która wymaga więcej czasu do zrozumienia i zaimplementowania. Dodatkowo sieci typu GAN są dość złożonymi modelami w porównaniu do sieci CNN, czy RNN, więc w kontekście generowania muzyki efekty można uzyskać znacznie szybciej implementując prostszą architekturę. Podsumowując, uważam, że przebieg pracy inżynierskiej był dość dobrze zaplanowany. Spełniłem założenia początkowe, czyli generowanie muzyki, a ewentualne trudności stanowią dobrą lekcję na przyszłość.

# 8 Bibliografia

# 9 Wykaz symboli i skrótów

| | |
|---|---|
| GAN | Generative Adversarial Network |
| LSTM | Long Short Term Memory |
| CNN | Convolutional Neural Network |
| RNN | Recurent Neural Network |
| AE | Auto Encoder |
| VAE | Variational Auto Encoder |
| GPU | Graphics Processing Unit |
| CPU | Central Processing Unit |
| CUDA | Compute Unified Device Architecture |
| MAESTRO | MIDI and Audio Edited for Synchronous TRacks and Organization |
| API | Application Programming Interface |
| GC | Google Colaboratory |

# 10    Spis rysunków

# 11  Załączniki

W skład załączonej płyty CD wchodzi kopia tej pracy inżynierskiej, program, który służył do generowania muzyki, skrypty do generowania zbioru uczącego, niezbędne oprogramowanie, przykłady wygenerowanych obrazów i muzyki, wyniki eksperymentów i instrukcja.

## 11.1  Sieć neuronowa i dane uczące

W tym folderze znajduje się sieć neuronowa ucząca się na zamieszczonym zbiorze treningowym, pliki konfiguracyjne środowiska uruchomieniowego i niezbędne do treningu dane uczące. Dodatkowo zamieściłem zapisane wagi z ostatniego z eksperymentów, aby odtworzyć wynik tamtego eksperymentu bez przeprowadzania długotrwałego uczenia. Sposób instalacji środowiska i uruchomienia sieci neuronowej jest opisany w pliku README.txt w głównym katalogu płyty CD.

## 11.2  Skrypty do generowania danych uczących

W tym katalogu znajduje się fragment bazy plików midi (100 nagrań – około 9% całej bazy). Dodatkowo zamieszczone są skrypty generujące zbiór uczący. Proces generowania zbioru uczącego jest opisany w pliku README.txt w głównym katalogu płyty CD.

## 11.3  Przykłady i wyniki eksperymentów

W tym folderze znajdują się przykładowe obrazy z baz treningowych, pliki midi z bazy nagrań muzyki jazzowej i klasycznej oraz wyniki wszystkich czterech eksperymentów. Zamieszczone są funkcje kosztu, porównanie obrazów prawdziwych i fałszywych, animacje z procesu uczenia oraz najważniejsze, czyli powstałe nagrania MIDI.

## 11.4  Załączone oprogramowanie

Załączone oprogramowanie jest niezbędne do przeprowadzenia większości eksperymentów, a jego instalacja nie powinna sprawiać problemów, w skład załączonego oprogramowania wchodzą:

- Midi2Mtx – program służący do przetwarzania plików midi do formy tekstowej – plików MTX

- Mtx2Midi – program służący do przetwarzania plików MTX w formie tekstowej do plików MIDI

- MidiPlayer5 – odtwarzacz MIDI